\documentclass[journal]{IEEEtran}
\usepackage{graphicx}
\usepackage{amsmath,amssymb,dsfont,bbm,epstopdf,pgfplots,mathtools,enumitem,mathrsfs, bbm, algpseudocode , graphicx, dblfloatfix, booktabs, physics, algorithm,algcompatible, derivative, physics} 
\usepackage[normalem]{ulem}
\usepackage{color}
\usepackage{cite}
\usepackage{graphics}
\usepackage{tikz}
\usepackage{pgfplots}
\usepackage{subcaption}
\usepackage[utf8]{inputenc} 
\usepackage[T1]{fontenc}
\usepackage[left=0.6in,right=0.6in,top=0.71in, bottom = 0.94in]{geometry}
\usetikzlibrary{shapes, arrows, decorations.markings, arrows.meta}
\usetikzlibrary{patterns} 
\allowdisplaybreaks
\graphicspath{{.}{./Figures/}} 
\allowdisplaybreaks[4] 

\newtheorem{proposition}{Proposition}

\newcommand{\vect}[1]{\boldsymbol{#1}}
\newcommand{\vectsf}[1]{\boldsymbol {\mathsf {#1}}}

\definecolor{ForestGreen}{rgb}{0.0, 0.5, 0.0}

\renewcommand{\P}{\mathsf{P}}

\newcommand{\C}{\mathsf{C}}

\renewcommand{\P}{\mathsf{P}}

\begin{document}
\title{ A Broadcast Channel Framework for MIMO-OFDM Integrated Sensing and Communication}
\author{\IEEEauthorblockN{\hspace{-0.5cm} Homa Nikbakht, Husheng Li, \IEEEmembership{Senior Member,~IEEE}, Zhu Han, \IEEEmembership{Fellow,~IEEE}, and H.~Vincent Poor, \IEEEmembership{Life Fellow,~IEEE}}

	\thanks{H.~Nikbakht and H.~V.~Poor are with the Department of Electrical and Computer Engineering, Princeton University, NJ, USA (email: \{homa, poor\}@princeton.edu)).  
H.~Li is with the School of Aeronautics and Astronautics, and the Elmore
Family School of Electrical and Computer Engineering, Purdue University,
USA (email: husheng@purdue.edu).  Z.~Han is with the Department of Electrical Engineering, University of
Houston (email: hanzhu22@gmail.com).  
}}

\maketitle

\begin{abstract}
Integrated sensing and communication (ISAC) is expected to be one of the major features of 6G wireless networks. In an ISAC system, communications and sensing functionalities are jointly performed using the same waveform, frequency band and hardware, thereby enabling  various use cases such as  in cyber physical systems, digital twin and smart cities.  A major challenge to the design and analysis of ISAC  is a unified framework that incorporates the two distinct functions.  By viewing ISAC as a type of broadcast channel, in this paper, we propose a unified ISAC framework  in which communication and sensing signals are broadcast to the actual  communication users and virtual sensing users. This framework allows the application of existing multiplexing schemes, such as dirty paper coding (DPC) and frequency division multiplexing (FDM) that have been intensively studied in data communications and information theory. Within this framework, we propose different superposition coding schemes,
for  cases when the sensing waveform is known or unknown
to the communication receiver. 
We propose the waveform optimization algorithms in a multiple-input multiple-output (MIMO) setting accounting for the effects of clutter and Doppler shift. 
 The proposed framework is numerically evaluated for different schemes under various sensing and communications performance metrics. 
\end{abstract}

\section{Introduction}\label{sec:intro}
Integrated sensing and communication (ISAC)  is envisioned to be a key enabling technology in 6G cellular networks \cite{FanLiu2020, Visa2024, Wei2025}. ISAC designs offer various use cases in cyber physical systems such as vehicular ad hoc networks and urban air mobility. In such applications,  each mobile node needs to simultaneously communicate with neighbors and sense the environment. The integration of  the two closely related but significantly different functions in the same waveform, hardware and software, introduces new design challenges.  One of the main challenges in ISAC system design is the need for a unified theoretical framework. Historically, communications and radar sensing are based on information theory and detection/estimation theory, respectively. Although the two theories have many overlaps, (for example, using the information-theoretic metrics for analyzing the detection/estimation performance, such as in the Stein's Lemma \cite{Cover2006}) they differ in their performance metrics and analytical tools. Communication systems typically  focus on metrics such as channel capacity and rely on techniques like random coding, whereas, in radar sensing systems the focus is on metrics such as the minimum mean square error (MMSE) and the use of  the Cram\'er-Rao bound. 

One effective unified ISAC framework can be built by leveraging work on broadcast channels (BCs) in information theory and downlink cellular networks. Comparing ISAC to downlink broadcast communications, sensing can be considered as a `virtual' user sharing the bandwidth with the actual communication user. 
The BCs framework offers various advantages for ISAC system design:
(a) The extensive body of research on BCs, particularly in the multiple-input multiple-output (MIMO) setting, provides valuable insights when applied to ISAC. In particular, the understanding of  how information is layered and superimposed for delivery to multiple destinations is essential in ISAC designs; and  
(b) Concrete algorithms for data multiplexing developed for  BCs can be applied to ISAC. For example,  advanced schemes such as dirty paper coding (DPC) \cite{Costa1983} and linear precoding \cite{Heath2019} can be employed to effectively mitigate interference from sensing signals on communication signals.

\begin{figure}
  \centering
  \includegraphics[width=0.4\textwidth]{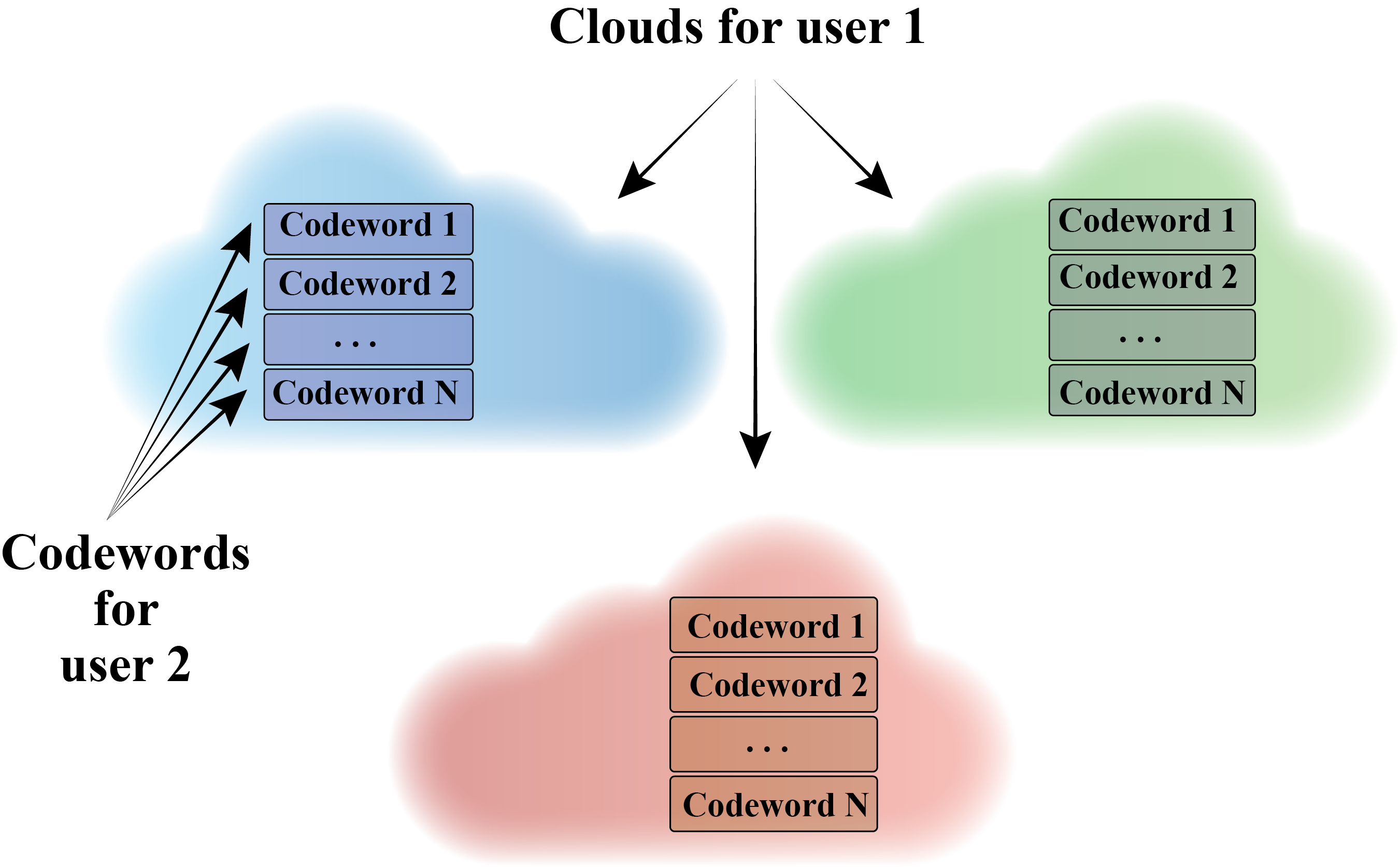}
  \caption{Superposition coding.}\label{fig:cloud}
  \vspace{-0.3cm}
\end{figure}
Motivated by the BC framework, in this paper, we propose a unified framework for ISAC that bridges sensing waveform design and communication channel coding.
As illustrated in Fig. \ref{fig:cloud}, broadcast channel coding can be implemented using superposition coding \cite{WangLele2013,Vanka2012,LeleWang2020,RZhang2010}, in which the messages of user~$1$ can be represented by the `clouds' while the messages of user~$2$ are encoded into each cloud. For decoding, user~$1$ simply determines which cloud is transmitted, while user~$2$ first determines the cloud and then estimates the codeword within the cloud, in an onion-peeling manner. 
In the context of ISAC, sensing and communications  can be alternatively modeled as users~$1$ and $2$ which leads to the following layered signaling schemes: Communication-centric (CC) and sensing-centric (SC) schemes.  In the CC scheme, the communication signal is laid at the top layer and  has the higher
priority,  whereas in the SC scheme, sensing takes the top layer and is accordingly prioritized. 
The main focus of this paper is on the CC scheme. In this case, sensing is considered as the bottom user in the layered structure of superposition coding.  A sensing waveform thus is generated according to a certain design metric, which plays the role of the cloud in the superposition coding. The main contributions of this work are:
\begin{itemize}
\item By viewing ISAC as a type of BC, we propose a unified ISAC framework  in which communication and sensing signals are broadcast to the actual  communication users and virtual sensing users. 

\item To further adapt this unified framework to the environment, 
we consider two practical scenarios: waveform available at receiver (WAR); and waveform unavailable at receiver (WUR). Both scenarios are commonly encountered in real-world applications. Specifically, in the WAR scenario, the communication receiver has full knowledge of the sensing waveform (cloud) employed  by the ISAC transceiver. This is reasonable when the environment changes slowly, such that the optimal waveform does not change rapidly.
In the WUR scenario,  the communication receiver does not have full knowledge of  the exact sensing waveform (cloud), and it only knows the set of possible waveforms. 
\item For both WAR and WUR scenarios, we study three superposition coding schemes: DPC, spectrum spreading, and linear transformation. 

\item Under the CC scheme, we consider three design metrics: 1) maximizing signal to clutter plus noise (SCNR) ratio, 2) minimizing the ambiguity function (AF) sidelobes under the effect of Doppler shift, and 3) minimizing the expected integrated sidelobe level (ISL). We propose concrete algorithms for waveform synthesis for each design. 

\item Through our numerical analysis, we compare the performance of the proposed  superposition coding schemes under various design criteria.  We also study the feasible ISAC performance region within a two-layer superposition coding signaling framework consisting of both CC and SC schemes. The established ISAC performance region allows  studying the sensing and communication trade-offs under various sensing and communication performance metrics.
\end{itemize}




The remainder of the paper is organized as follows. Related works are discussed in Section~\ref{sec:related}. We present the system model in Section~\ref{sec:model}. We introduce different communication and sensing performance metrics in Section~\ref{sec:metric}.  The feasible ISAC performance region is established in Section~\ref{sec:region}.  Superposition coding schemes for WAR and WUR are discussed in Section \ref{sec:superposition}. Concrete  sensing waveform optimization algorithms are provided in Section~\ref{sec:design}. Finally, the numerical results and conclusions are provided in Sections \ref{sec:numerical} and \ref{sec:conclusions}, respectively. 


\section{Related Works} \label{sec:related}
Various ISAC designs have been studied in the literature \cite{LiSPAWC2022, Mittelbach2025, Ahmadi2024, Liu2025, Nikbakht2024, Zhao2025, Li2023,Yang2025, Wei2024, Li2025, HomaAsilomar2024, Zhang2025, Bian2025}. Notable examples include unsourced random access \cite{Ahmadi2024}, multiple access \cite{Liu2025}, and finite blocklength \cite{Nikbakht2024, Zhao2025} fundamental limits of ISAC, optimal precoding and waveform designs for MIMO-OFDM-based ISAC systems \cite{Li2023,Yang2025, Wei2024, Li2025}, and the application of deep learning to enhance the performance of such systems \cite{HomaAsilomar2024, Zhang2025, Bian2025}. Comprehensive surveys on ISAC can be found in \cite{FanLiu2020, Visa2024, Wei2025}. For data communications, BC has been intensively studied in early 2000s. While the BC channel capacity region has been identified for degraded channels \cite{Cover2006}, it is still an open problem for the MIMO case. A major breakthrough is the introduction of DPC \cite{Costa1983}, disclosed by \cite{Caire2003}. Based on DPC, the duality between multiple-access channel (MAC) and BC is identified in \cite{Jindal2004}. 
Superposition coding has been studied for both broadcast and interference channels, mainly from the viewpoint of information theory. Different types of theoretical superposition coding schemes have been compared in \cite{WangLele2013}. More practical coding schemes have been proposed in \cite{Vanka2012}, and a survey on superposition coding can be found in \cite{RZhang2010}.

In the pioneering study by D. Bliss \cite{Paul2017_2}, the integration of communications and sensing is studied in which sensing is also considered as a user characterized by its information rate. Similar ideas are considered by X.~Mu et al.  in \cite{XidongMu2023}, where successive interference cancellation is employed for mitigating the interference between sensing and communication signals. Different from the multiple access channels considered by these studies, this paper focuses on the broadcast of both communications and sensing from a single transmitter, thus requiring substantially different coding methodologies. Compared to the single-input single-output (SISO) setting investigated in the conference version of this study \cite{LiGlobecom20231,LiGlobecom20232}, in this paper, we consider a MIMO-OFDM setting and expand our waveform design analysis to take into account the effect of clutter and Doppler shift.  

\section{System Setup and Problem Formulation} \label{sec:model}
In this section, we introduce the setup of our ISAC system and the formulation of our problem.
\vspace{-0.2cm} 
\subsection{Transmit Signal}
Consider a monostatic ISAC setting  with a transceiver of transmit power $\P_t$ and bandwidth $\mathsf B$, a communication receiver and a radar target.  Denote by $N$, the number of transmit and receive antennas at the ISAC transceiver and by  $N_c$  the number of antennas at the communication receiver.

 We employ orthogonal frequency-division multiplexing (OFDM) signaling with $M_c$ subcarriers and frequency spacing $\Delta f: = \frac{1}{T_d}$ where $T_d$ is the OFDM symbol duration.  Let $L$ be the frame length of one coherent processing interval (CPI). For the $\ell$-th symbol with $\ell \in \{0, \ldots, L-1\}$, we denote the dual-function baseband signal on the $m$-the subcarrier as $\vect x_{\ell,m}\in \mathbb C^{N}$.

 At the ISAC transmitter, the $M_c$ $N$-dimensional frequency baseband signals are rearranged into $N$ $M_c$-dimensional vectors collected from different antennas. Then $M_c$-point inverse discrete Fourier transform (IDFT) processors are employed to transform frequency-domain signals into time-domain ones. To avoid inter-symbol interference (ISI), an $M_{\text{cp}}$-point cyclic prefix (CP) of duration $T_{CP}$ is inserted. After converting to analog, the  baseband signal is given by  
 \begin{eqnarray}
\vect x(t)=\sum_{\ell=0}^{L-1}\sum_{m=0}^{M_c-1} \vect x_{\ell,m} e^{j2\pi m\Delta f t}\; \text{rect} \left (\frac{t-LT_c}{T_c}\right) ,
\end{eqnarray}
where $T_c:= T_d + T_{CP}$ is the total symbol duration and $\text{rect}(t/T_c)$ denotes a rectangular pulse of duration $T_c$. In our analysis, 
\begin{equation}
  \vect x(t) = f_{\text{sp}} (\vect x^{(s)}(t), \vect x^{(c)}(t)),
\end{equation}
where $\vect x^{(s)}(t): = \{ x_1^{(s)}(t), \ldots,  x_{N}^{(s)}(t)\}$ and $\vect x^{(c)}(t): = \{ x_1^{(c)}(t), \ldots,  x_{N}^{(c)}(t)\}$ are the corresponding sensing and communication baseband signals, respectively, and $f_{\text{sp}} (\cdot, \cdot)$  is the superposition function which depends on the type of superposition coding scheme that we will discuss in Section~\ref{sec:superposition}.  


The baseband analog signal is then up-converted to the radio frequency (RF) domain using $N$ RF-chains resulting in the following signal 
\begin{equation}
\tilde{\vect x}_{\text{RF}}(t)= \vect x(t) e^{j2\pi f_c t},
\end{equation}
where $f_c$ is the carrier frequency. The signal is then emitted through the antennas. 


\subsection{Received Echos}
The transmitted signal $\tilde{\vect x}_{\text{RF}}(t)$ is reflected by the target and the echo signal is collected by the ISAC receiver. Assume that there are $\mathcal G$  major
clutterers in space, and that the radar cross section (RCS) of the target is constant during the total OFDM transmission $LT_c$.  After down-conversion, the baseband echo signal received by the $N$ receiving antennas at the BS is given by 
\begin{IEEEeqnarray}{rCl}
\lefteqn{\vect y^{(r)}(t)}\notag \\ 
 &=& \alpha \vect a_r(\theta_0) \vect a_t^H(\theta_0) \vect x(t-\tau) e^{-j2\pi f_c \tau} e^{j2\pi f_dt} \notag \\
 && + \sum_{g = 1}^{\mathcal G} \alpha_g \vect a_r(\theta_g) \vect a_t^H(\theta_g) \vect x(t-\tau_g) e^{-j2\pi f_c \tau_g} e^{j2\pi f_{d,g}t} \notag \\
 && \hspace{0.5cm}+ \vect z_s(t) \IEEEeqnarraynumspace\\
& = &   \vectsf H_0 \sum_{\ell=0}^{L-1}\sum_{m=0}^{M_c-1} \vect x_{\ell,m} e^{j2\pi m\Delta f (t-\tau)}\;  \text{rect} \left (\frac{t- \tau-mT_c}{T_c}\right) e^{j2\pi f_dt} \notag \\
&& + \sum_{g = 1}^{\mathcal G}\vectsf H_{g,0} \sum_{\ell=0}^{L-1}\sum_{m=0}^{M_c-1} \vect x_{\ell,m} e^{j2\pi m\Delta f (t-\tau_g)}\; \notag \\
&& \hspace{0.5cm}\times  \text{rect} \left (\frac{t- \tau_g-mT_c}{T_c}\right) e^{j2\pi f_{d,g}t} \notag \\
&& \hspace{1cm}+ \vect z_r(t), \label{eq:yrt}
\end{IEEEeqnarray}
where $ \alpha := \sqrt{\frac{\delta_{\text{RCS}}\lambda^2}{(4\pi)^3 R_0^4}}$ denotes the attenuation factor with $\delta_{\text{RCS}}$ the RCS of the target, $\lambda = c/f_c$ the wavelength, $c$ the speed of light, and $R_0$ the target range. $ \vectsf H_0:= \alpha e^{-j2\pi f_c \tau} \vect a_r(\theta_0) \vect a_t^H(\theta_0)$, $\tau:=\frac{2R_0}{c}$ is the round-trip delay, $f_d:=\frac{2\nu_0f_c}{c}$ is the Doppler shift with $\nu_0$ being the target velocity, $\theta_0$ is the angle of both arrival and departure of the target, and 
\begin{IEEEeqnarray}{rCl}
\vect a_t(\theta_0) &:=& \left [1, e^{j 2\pi \sin (\theta_0) d_t/\lambda}, \ldots, e^{j 2\pi (N -1)\sin (\theta_0) d_t/\lambda}\right]^T, \IEEEeqnarraynumspace \\
\vect a_r(\theta_0) &:=& \left [1, e^{j 2\pi \sin (\theta_0) d_r/\lambda}, \ldots, e^{j 2\pi (N -1)\sin (\theta_0) d_r/\lambda}\right]^T, \IEEEeqnarraynumspace
\end{IEEEeqnarray} 
denote the transmit and receive steering vectors, respectively, with $d_t$ and $d_r$ as the corresponding transmit and receive antenna spacings. The same definitions are applied to the corresponding clutterers' parameters. The vector $\vect z_r(t)$ denotes additive white Gaussian noise (AWGN). 
 
 The baseband signals in \eqref{eq:yrt} are first sampled using analog-to-digital converters (ADCs) with sampling rate $f_s = M_c \Delta f$ which results in the following discrete-time radar echo signal: 
 \begin{IEEEeqnarray}{rCl}
\lefteqn{\tilde{\vect y}^{(r)}_{k,\ell}} \notag \\
  &=& h_0 \sum_{m=0}^{M_c-1} \vect x_{\ell,m} e^{j2\pi m\frac{k}{M_c}}e^{-j2\pi m \Delta f \tau} e^{j2\pi f_d T_c \left ( \frac{k}{M_c} + \ell \right) } \notag \\
 &&+   \sum_{g = 1}^{\mathcal G}  h_g \sum_{m=0}^{M_c-1} \vect x_{\ell,m} e^{j2\pi m\frac{k}{M_c}}e^{-j2\pi m \Delta f \tau_g} e^{j2\pi f_{d,g} T_c \left (  \frac{k}{M_c} + \ell \right) } \notag \\
  &&\hspace{0.5cm} + \tilde{\vect z}_{k,\ell}
 \end{IEEEeqnarray}
 where  $\tilde{\vect y}_{k,\ell}$ denotes the $k$-the sample of the $\ell$-th OFDM symbol in  $\vect y_r(t)$ and $\tilde{\vect z}_{k,\ell} = \vect z_r(\frac{kT_c}{M_c} + \ell T_c)$. The exponential term $e^{j2\pi m\frac{k}{M_c}}$ represents the effect of OFDM modulation, $e^{-j2\pi m \Delta f \tau}$ is the delay-induced phase shift over the OFDM subcarriers, $e^{j2\pi \frac{k}{M_c}f_d T_c}$ is the Doppler-induced shift in the fast-time domain, and $e^{j2\pi \ell f_d T_c}$ is the Doppler-induced shift in the slow-time domain. To avoid inter-carrier interference (ICI), we choose $\Delta f$ to be larger than the Doppler shifts, i.e., $f_d T_c = f_d/\Delta f \ll 1$ and $ f_{d,g}/\Delta f \ll 1$ for each $g \in \{1, \ldots, \mathcal G\}$. We thus omit the effect of Doppler-induced shifts $e^{j2\pi \frac{k}{M_c}f_d T_c}$ and $e^{j2\pi \frac{k}{M_c}f_{d,g} T_c}$ for $g \in \{1, \ldots, \mathcal G\}$. Next, we perform a discrete Fourier transform (DFT) over each OFDM symbol and convert the discrete-time signal $ \tilde{\vect y}_{k,\ell}$ into the following frequency domain signal:
 \begin{IEEEeqnarray}{rCl}
 \vect y^{(r)}_{\ell,m} &=&  \vectsf H_0\vect x_{\ell,m} e^{-j2\pi m \Delta f \tau} e^{j2\pi \ell f_d T_c} \notag \\
 &&+ \sum_{g = 1}^{\mathcal G} \vectsf H_{g,0} \vect x_{\ell,m} e^{-j2\pi m \Delta f \tau_g} e^{j2\pi \ell f_{d,g} T_c} + \vect z_{\ell,m}, \label{eq:yml} \\
 & = & \vectsf H_{t,\ell,m} \vect x_{\ell,m} + \sum_{g = 1}^{\mathcal G} \vectsf H_{g,\ell,m} \vect x_{\ell,m} + \vect z_{\ell,m}, \label{eq:10n}
 \end{IEEEeqnarray}
  where $\vect z_{\ell,m}$ represents the Fourier transform of the noise  $\tilde{\vect z}_{k,\ell}$, and 
  \begin{IEEEeqnarray}{rCl}
  \vectsf H_{t,\ell,m} &:=&\vectsf H_0 e^{j2\pi (\ell f_d T_c - m \Delta f \tau)},  \\
  \vectsf H_{g,\ell,m} &:=&  \vectsf H_g  e^{j2\pi \left (\ell f_{d,g} T_c - m \Delta f \tau_g\right)},
  \end{IEEEeqnarray}
 for $g \in \{1, \ldots, \mathcal G\}$.

\subsection{Received Signal at the Communication Receiver}
We assume that the communication channel experiences frequency selective fading. The received signal at the communication user is down-converted to baseband, followed by analog- to-digital conversion, CP removal, and a DFT. The received signal on the $m$-th subcarrier of the $\ell$-th symbol is given by
\begin{IEEEeqnarray}{rCl}
\vect y_{\ell,m}^{(c)} = \vect x_{\ell,m} \vectsf H_{c,\ell,m}   + \vect z_{\ell,m}^{(c)},
\end{IEEEeqnarray}
where $ \vectsf H_{c,\ell,m} \in \mathbb C^{N\times N_c}$ is the frequency domain  channel response and $ \vect z_{\ell,m}^{(c)} \in \mathbb C^{1\times N_c}$ is AWGN whose  entries are i.i.d. $\mathcal N(0,\sigma_c^2)$.


\section{Performance and Design Metrics} \label{sec:metric}
In this section, we introduce the performance and design metrics for sensing and communication tasks. 

\subsection{Communication Performance Metric: Capacity}
The communication channel capacity is given by 
\begin{equation} \label{eq:capacity}
\C = \frac{1}{M_c L} \sum_{\ell = 0}^{L-1}  \sum_{m = 0}^{M_c-1}\log \det \left (\textbf{I}_{N_c} + \frac{1}{\sigma_c^2} \vectsf H_{c,\ell,m} \vectsf R_{\ell,m} \vectsf H_{c,\ell,m}^H \right),
\end{equation}
where $\vectsf R_{\ell,m}:= \mathbb E [\vect x_{\ell,m} \vect x_{\ell,m}^H]$ is the transmit covariance matrix which is subject to the following power constraint: 
\begin{IEEEeqnarray}{rCl}
\sum_{\ell = 0}^{L-1} \sum_{m = 0}^{M_c-1}  \text{Tr}(\vectsf R_{\ell,m}) \le \P_t.
\end{IEEEeqnarray} 
\subsection{Sensing Performance Metrics}
\subsubsection{Ambiguity Function (AF)}
A useful performance characterization for radar sensing is the AF proposed by Woodward \cite{Woodward1980,He2012}. The AF represents the time-frequency composite auto-correlation function of the transmitted signal and is given by
\begin{IEEEeqnarray}{rCl}
\chi(\tau, f_d)=\int_{-\infty}^\infty x_0(t)x_0^*(t-\tau)e^{-j2\pi f_dt}dt,
\end{IEEEeqnarray}
where $x_0(t):= \vect a_t^H(\theta_0) \vect x(t)$ is the OFDM signal after beamforming to the angle $\theta_0$. Assuming that the round-trip delay $\tau$ is smaller than the CP duration, after some simplifications the discrete periodic AF function can be formulated as
\begin{IEEEeqnarray}{rCl} \label{eq:AFd}
\chi(d, \nu)&=&
\sum_{\ell = 0}^{L-1} \sum_{m=0}^{M_c-1} \tilde x_{\ell,m} \tilde x^*_{\ell-d,m} e^{\frac{j2\pi (\ell -d)\nu}{L}}, 
\end{IEEEeqnarray}
where $\tilde x_{\ell,m} := \vect a_t^H(\theta_0) \vect x_{\ell,m} $, and  $d:= \tau M_c \Delta_f$ and $\nu = f_d L T_c$ are the indices of the range and Doppler bins, respectively. 

\subsubsection{Integrated Sidelobe Level (ISL)}
For sensing purposes, it is desirable to have a narrow mainlobe peak at $(d = 0, \nu = 0)$ and low sidelobes for other points in the $d$-$\nu$ plane. A dual-function waveform ISAC system is also used for conveying information and thus  tends to have a strong random component. This will  lead to higher sidelobe levels compared to deterministic waveforms like Zadoff–Chu sequences. Thus, an important consideration in ISAC waveform design is how to control the sidelobe levels. 
%
Therefore,  the ISL \cite{He2012} is considered to be an important sensing design metric and is given by
\begin{IEEEeqnarray}{rCl}\label{eq:ISL}
\xi_{\text{ISL}}(\vect x)
 &=& \sum_{d \in \mathcal A} \sum_{\nu \in \mathcal B} | \chi(d,\nu) |^2 - | \chi(0,0) |^2,
\end{IEEEeqnarray}
where $\mathcal A$ and $\mathcal B$ are the time delay and the Doppler frequency sets of interest, respectively. 

\subsubsection{Signal-to-Clutter-plus-Noise Ratio (SCNR)} 
At a given frequency $w$, let $\vectsf G_{\ell,m}(w)$ to be the power spectral density associated with the noise process $\vect z_{\ell,m}$ in \eqref{eq:10n} and $\vectsf S_{F,\ell,m}(w)$ be the total clutter power given by 
\begin{equation}
\vectsf S_{F,\ell,m}(w) = \sum_{g = 1}^{\mathcal G} \vectsf H_{g,\ell,m}(w) \vect X_{\ell,m}(e^{jw})\vect X^*_{\ell,m}(e^{jw}).
\end{equation} 
The output SCNR at the decision instant $t = t_0$ is thus given by 
\begin{IEEEeqnarray}{rCl}\label{eq:scnr}
\gamma_{\text{SCNR}} (t_0)&:=&  \sum_{\ell = 0}^{L-1}  \sum_{m = 0}^{M_c-1} \tilde \gamma_{\text{SCNR}, \ell,m} (t_0),\notag \\ \IEEEeqnarraynumspace
\end{IEEEeqnarray}
where 
\begin{IEEEeqnarray}{rCl}\label{eq:scnr2}
\tilde \gamma_{\text{SCNR}, \ell,m} (t_0)
&:=&  \frac{\left | \frac{1}{2\pi} \int_{-\infty}^{\infty} \vectsf H_{t,\ell,m}(e^{jw}) \vect X_{\ell,m}(e^{jw}) e^{jwt_0}dw\right |^2}{\frac{1}{2\pi} \int_{-\infty}^{\infty} \vectsf S_{\ell,m} (w)dw  } \notag \\
\end{IEEEeqnarray}
with
\begin{equation}\label{eq:S}
\vectsf S_{\ell,m} (w):= \vectsf S_{F,\ell,m}(w) + \vectsf G_{\ell,m}(w),
\end{equation}
for each $m \in \{0, \ldots, M_c -1\}$ and each $\ell \in \{0,\ldots, L-1\}$.

\section{Broadcast Channel Framework of ISAC} \label{sec:region}
\begin{figure}
  \centering
  \includegraphics[width=0.35\textwidth]{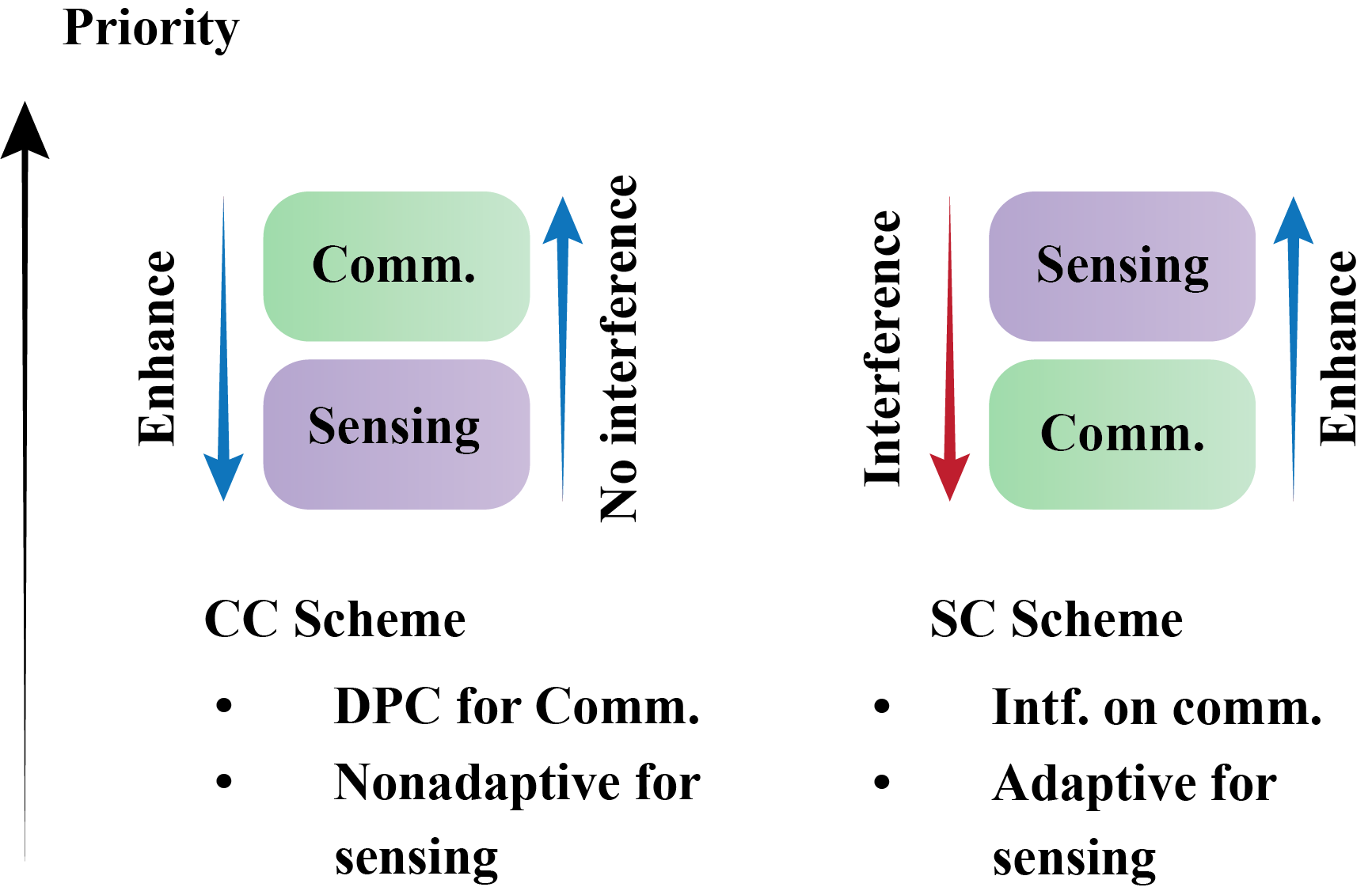}
  \caption{Two layered signaling schemes.}\label{fig:two_schemes}
  \vspace{-0.4cm}
\end{figure}

As illustrated in Fig. \ref{fig:cloud}, to specify  the two-user broadcast channel framework using superposition coding \cite{WangLele2013,Vanka2012,LeleWang2020,RZhang2010}, the messages of user~$1$ can be represented by the `clouds' while the messages of user~$2$ are encoded into each cloud. For decoding, user~$1$ simply determines which cloud is transmitted, while user~$2$ first determines the cloud and then estimates the codeword within the cloud, in an onion-peeling manner. 
In the context of ISAC, sensing and communications  can be alternatively modeled as users~$1$ and $2$ which leads to the following layered signaling schemes:

\begin{itemize}
\item \emph{Communication-centric (CC) scheme}: In this scheme, communication is laid at the top layer and thus has the higher priority. The sensing waveform $\vect x^{(s)}$ will be synthesized first, independent of the realization of communication messages (but could be dependent on the corresponding statistics). Then, the communication signal $\vect x^{(c)}$ is formed as  a function of the sensing waveform and the communication message through different superposition coding techniques explained in Section~\ref{sec:superposition}. 

\item \emph{Sensing-centric (SC) scheme}: In this scheme, sensing is laid at the top layer and thus has the higher priority (see Fig.~\ref{fig:two_schemes}). The communication signal $\vect x^{(c)}$ is generated first based on some design criteria. Typically, under this scheme, the communication signal is generated subject to the interference of the sensing signal. Then, the sensing signal is superposed on the top of  the realization of communication signal using a selected superposition technique from Section~\ref{sec:superposition}. 
\end{itemize}

\begin{figure}
  \centering
  \includegraphics[width=0.32\textwidth]{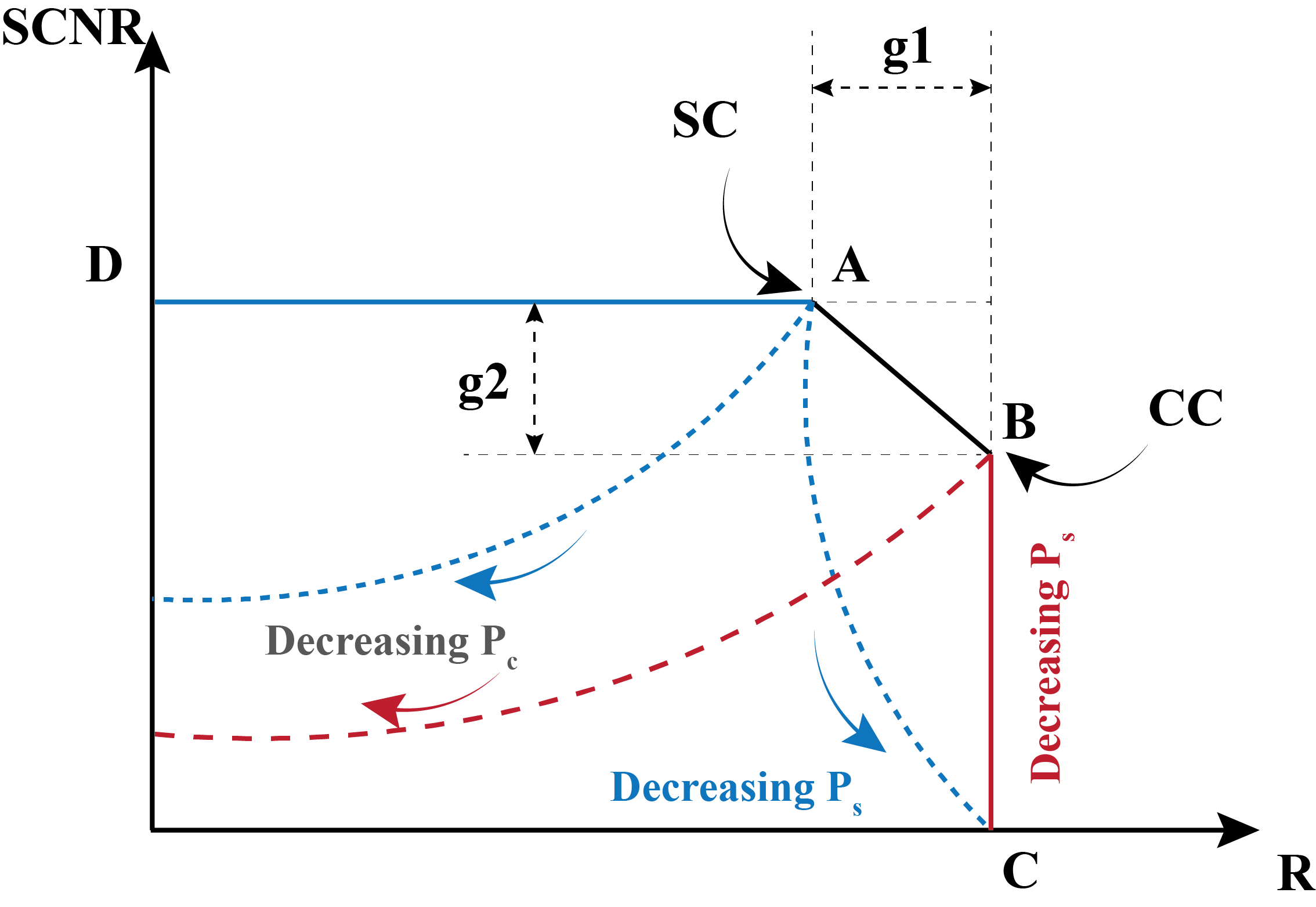}
  \caption{Feasible performance region for ISAC.}\label{fig:region2}
  \vspace{-0.4cm}
\end{figure}

Given the above two layered signaling schemes, one can obtain feasible performance regions for ISAC depending on the desired performance metrics. For example, consider SCNR and data rate $R$ as the sensing and communication  performance metrics, respectively.  We allocate a portion $\P_c$ of the total transmit power $\P_t$ to  the communication task and the remainder $\P_s$  to the sensing task with $\P_c + \P_s = \P_t$. A schematic feasible ISAC region is illustrated in Fig. \ref{fig:region2}. In this region, we first fix the performance points for the CC (point B) and SC (point A) schemes.
 Since in the SC scheme,  communication experiences interference from the sensing signal, while the sensing waveform is optimized with respect to the generated communication signals, the performance point of the SC scheme is to the upper left of that of the CC scheme. 
The horizontal boundary $AD$ is obtained by decreasing the data rate $R$ without changing $\P_s$ and $\P_c$. The vertical boundary $BC$ is obtained by decreasing $\P_s$ while keeping $\P_c$ constant. The boundary $AB$ is then obtained from time-sharing between SC and CC. The gap $g1$ is the gap in $R$ that appears in cases where the sensing signal is considered as interference in the SC scenario. The gap $g2$ is the gap in MMSE due to the waveform adaptation.  The trajectories of decreasing $\P_c$ or $\P_s$ from $A$, and decreasing $\P_c$ from $B$, are all plotted in dashed lines  and can be shown to be within the region. 

The main focus of this paper is on the CC scheme. In this case, sensing is considered as the bottom user in the layered structure of superposition coding.  A sensing waveform thus is generated according to a certain criterion of sensing, which plays the role of the cloud in superposition coding. Throughout this work, we denote the optimum sensing waveform by $\vect x^{(s)}_{\text{opt}}$.  In the following Section~\ref{sec:superposition}, we discuss various superposition coding schemes. Then, in Section~\ref{sec:design}, we first optimize the sensing waveform based on various design metrics introduced in Section~\ref{sec:metric}. The final waveform is then formed by superposing the communication signal on the top of the sensing signal using techniques that we explain in Section~\ref{sec:superposition}. In Section~\ref{sec:numerical}, we briefly mention examples of the SC scheme to investigate the ISAC feasible region through numerical analysis.  

\section{Superposition Schemes}\label{sec:superposition}
In this section, we discuss different superposition coding schemes for the CC case. As noted in Section~\ref{sec:intro}, we consider two practical scenarios: waveform available at receiver; and waveform unavailable at receiver. In the WAR scenario, the communication receiver has full knowledge of the sensing waveform (cloud) employed  by the ISAC transceiver, i.e., $\vect x^{(s)}_{\text{opt}}$. This is reasonable when the environment changes slowly, such that the optimal waveform does not change rapidly.
In the WUR scenario,  the communication receiver does not know the exact sensing waveform (cloud), while it knows the set of possible waveforms. In the context of superposition coding, the communication receiver does not know the selected codeword of the virtual sensing user, while it knows the corresponding codebook. 
In the following sections, we explain superposition schemes corresponding to each scenario. 
\subsection{Waveform Available at Receiver (WAR) Case} \label{sec:war}
In this section, we consider the WAR case where the  waveform is optimized corresponding to the environment and the communication receiver is informed about the optimum sensing  waveform $\vect x_{\text{opt}}^{(s)}$. In the context of superposition coding, this means that the identity of the cloud is known at the communication receiver  which significantly reduces the decoding complexity. In the following we describe several possible superposition coding  schemes. 

\subsubsection{Dirty-Paper Coding (DPC) Scheme}\label{sec:DPC}
If interference is known at the transmitter before communication starts, the transmitter can mitigate this interference through DPC \cite{Costa1983}. In the ISAC setting, we use DPC to precancel the interference of the optimum sensing signal $\vect x^{(s)}_{\text{opt}}$  from the corresponding communication signal $\vect x^{(c)}$. Specifically, the final transmit signal is given by 
\begin{equation}
\vect x= \vect x_{\text{opt}}^{(s)} + \vect x^{(c)},
\end{equation}
where $\vect x^{(c)}$ is determined by both the communication message and the sensing signal. 

To realize the DPC scheme, we employ the lattice encoding and decoding approach proposed in \cite{Hindy2016} for MIMO settings. We start by reviewing some of the main properties of lattice codes. 

An $n$-dimensional  lattice code is defined by 
\begin{IEEEeqnarray}{rCl}
\Lambda:= \{\vect \lambda = \vectsf G\cdot  \vect i: \vect i \in \mathbb Z^{n}\},
\end{IEEEeqnarray}
where $\vectsf G$ is the $n \times n$ generator matrix. The \emph{fundamental} Voroni region $\mathcal V$ of the lattice is the set of points with minimum Euclidean distance to the origin, defined by 
\begin{equation}
\mathcal V = \{ \vect s: \arg \min_{\vect \lambda \in \Lambda} || \vect s - \vect \lambda|| = 0\}.
\end{equation}
The second moment and the normalized second moment of the lattice $\Lambda$ are defined as 
\begin{IEEEeqnarray}{rCl}
\sigma^2_{\Lambda} &=& \frac{1}{n \text{Vol}(\mathcal V)}\int_{\mathcal V} ||\vect s||^2 d\vect s, 
\end{IEEEeqnarray}
and
\begin{IEEEeqnarray}{rCl}
V(\Lambda) &=& \frac{\sigma^2_{\Lambda} }{(\text{Vol}(\mathcal V))^{\frac{2}{n}}},
\end{IEEEeqnarray}
respectively, where $\text{Vol}(\mathcal V)$ denotes the volume of $\mathcal V$ and  $V(\Lambda) \le \frac{1}{2\pi e}$ for any $n$. 
For the lattice $\Lambda$, the quantizer is defined as 
\begin{equation}
Q_{\mathcal V}(\vect s) = \vect \lambda, \; \text{if} \; \vect s \in \vect \lambda + \mathcal V,
\end{equation}
and the modulo-$\lambda$ operation corresponding to $\mathcal V$ is given by 
\begin{equation}
[\vect s] \mod \Lambda := \vect s - Q_{\mathcal V}(\vect s).
\end{equation}
The lattice $\Lambda$ is said to be nested in $\Lambda_1$ if $\Lambda \subseteq \Lambda_1$. 

We now explain the lattice encoding and decoding processes. In the encoding process, 
we first construct a communication codebook $\mathcal L_1 = \Lambda_1 \cap \mathcal V$ whose rate is given by 
\begin{equation}
R = \frac{1}{NM_cL} \log \frac{\text{Vol}(\mathcal V)}{\text{Vol}(\mathcal V_1)},
\end{equation}
and $\Lambda_1 \in \mathbb R^{NM_cL}$. 
We then use a common randomness vector $\vect d \in \mathbb R^{NM_cL}$ (dither) known at both ISAC transmitter and the communication receiver. The emitted lattice codeword thus is given by 
\begin{equation}
\vect x = [ \vect t - \vect d - \vect x^{(s)}_{\text{opt}}] \mod \Lambda = \vect t + \vect \lambda - \vect d - \vect x^{(s)}_{\text{opt}},
\end{equation} 
where $\vect \lambda = -Q_{\mathcal V}(\vect t - \vect d- \vect x^{(s)}_{\text{opt}})$, the dither $\vect d$ is uniform over $\mathcal V$, $\vect t$ is a lattice point drawn from $\Lambda_1$, and $\Lambda \in \mathbb R^{NM_cL}$. At the communication receiver, the received communication signal $\vect y^{(c)}:=  [(\vect y^{(c)}_0)^T, (\vect y^{(c)}_1)^T, \ldots,  (\vect y^{(c)}_{L-1})^T] \in  \mathbb C^{N_cLM_c}$ is first multiplied by a receiver matrix $\vectsf U \in \mathbb R^{NM_cL\times N_cLM_c}$ and the dither is removed by forming 
\begin{IEEEeqnarray}{rCl}
\tilde {\vect y}^{(c)} &=&  \vectsf U \vect y^{(c)} + \vect d \\
& = & \vect t + \vect \lambda + \vect w_z,
\end{IEEEeqnarray} 
where 
\begin{equation}
\vect w_z := (\vectsf U \vectsf H_c - \textbf I_{NM_cL})\left (\vect x + \vect x^{(s)}_{\text{opt}}\right) + \vectsf U \vect z^{(c)}.
\end{equation}

To decode $\vect t$, we use the minimum Euclidean distance lattice decoder where the true lattice point with high probability has the shortest Euclidean distance to the received point. This implies that 
\begin{equation}
\hat{\vect t} = [\arg \min_{\tilde{\vect t} \in \Lambda_1} ||\tilde {\vect y}^{(c)} - \tilde{\vect t}||^2] \mod \Lambda.
\end{equation}

\subsubsection{Linear Transformation Scheme} Under this scheme, the communication message is modulated by a matrix $\vectsf {X}_c \in \mathbb C^{M_c N\times M_c N}$ such that the $\ell$-th symbol of the signal sent by the ISAC transceiver is given by
\begin{eqnarray}\label{eq:linear}
\vect x_{\ell}= \vectsf {X}_c \vect x_{\text{opt}, \ell}^{(s)}, \label{eq:73}
\end{eqnarray}
where $\vect x_{\ell}$ and $\vect x_{\text{opt}, \ell}^{(s)}$
are the $\ell$-th symbol of  $\vect x$ and  $\vect x_{\text{opt}}^{(s)}$, respectively. 
The selection of $\vectsf X_c$ should make $\vect x_{\ell}$ within or close to the optimal waveform manifold. Hence, we set
\begin{equation}
\vectsf X_c = \vectsf F^{-1} \vectsf \Delta \vectsf F,
\end{equation}
where $\vectsf F$ is the $NM_c \times NM_c$ DFT matrix, $\vectsf F^{-1}$ is the IDFT matrix, and $\vectsf \Delta$ is a diagonal matrix given by  
\begin{equation}
\vectsf \Delta :=\text{diag} [\vectsf \lambda_0,\ldots,\vectsf \lambda_{M_c-1}] \in \mathbb C^{NM_c \times NM_c},
\end{equation}
where $\vectsf \lambda_m \in \mathbb C^{N \times N}$. Then, in \eqref{eq:73}, $\hat {\vect x}_{\text{opt}, \ell}^{(s)} = \vectsf F \vect x_{\text{opt}, \ell}^{(s)}$ are the subcarriers of the sensing waveform (in the terminology of OFDM). These subcarriers then are multiplied by $\{\vect \lambda_m \}_{m = 0}^{M_c-1}$.
 To assure no performance loss for sensing, we adopt phase shift keying (PSK) for subcarriers of significant power, such that the power spectrum of $\vect x^{(s)}_{\text{opt}}$ is unchanged. Specifically, for each $m \in \{0, \ldots, M_c-1\}$,
\begin{IEEEeqnarray}{rCl}
\vectsf \lambda_m=
\begin{cases}
\text{diag}[e^{j\theta_1}, \ldots,  e^{j\theta_N}] ,\quad & ||\hat {\vect x}_{\text{opt}, \ell,m}^{(s)}||^2\ge \bar{x},\\
\textbf I_{N} ,\quad & \text{o.w.}
\end{cases},
\end{IEEEeqnarray}
where $\bar{x}$ is a predetermined threshold, $\textbf I_{N}$ is an $N \times N$  identity matrix, and $\theta_i$s with $i \in \{1, \ldots, N\}$ are the angles selected by the PSK modulation. 
Then, the IDFT $\mathbf{F}^{-1}$ converts the frequency domain back to the time domain.

Now, the major challenge is to determine the order of PSK in each subcarrier. We consider the following two approaches:
\begin{itemize}
\item \emph{Fixed $K$-PSK approach}: We fix the order of the PSK at $K$. The advantage  of this approach is that the communication receiver does not need to know the sensing waveform for the information of modulation order $K$. However, the fixed modulation order $K$ may waste bandwidth at subcarriers with large power while incurring unreliability of transmission at subcarriers with small power allocation. 
\item \emph{Adaptive modulation approach}: To handle the disparity of power allocation, we can consider  an adaptive modulation approach with respect to the power spectrum and the channel gain. To this end, we consider the schemes of $K=2^k$-PSK, $k=1,2,...$. For each subcarrier $m$, when using the $2^k$-PSK, the error rate of each bit is denoted by $P_e^k(\P_{\ell,m})$, which is a function of the power $\P_{\ell,m}$ at subcarrier $m$ of the $\ell$-th symbol. The function for PSK error rate can be obtained by numerical integral (p.194, \cite{Proakis2007}). Then, the estimate is obtained as
\begin{equation}
k^*=\arg\max_{k}k\left(1- H\left(P_e^k\left(\P_{\ell,m}\right)\right)\right),
\end{equation}
which means to select the modulation scheme yielding the maximum reliable throughput. The disadvantage of the adaptive modulation is that, when the sensing waveform $\vectsf x_{\text{opt}}^{(s)}$ is unknown to the communication receiver, it needs to be estimated by the communication receiver in order to obtain the modulation order $K$. 
\end{itemize}


\begin{figure}
  \centering
  \includegraphics[width=0.24\textwidth]{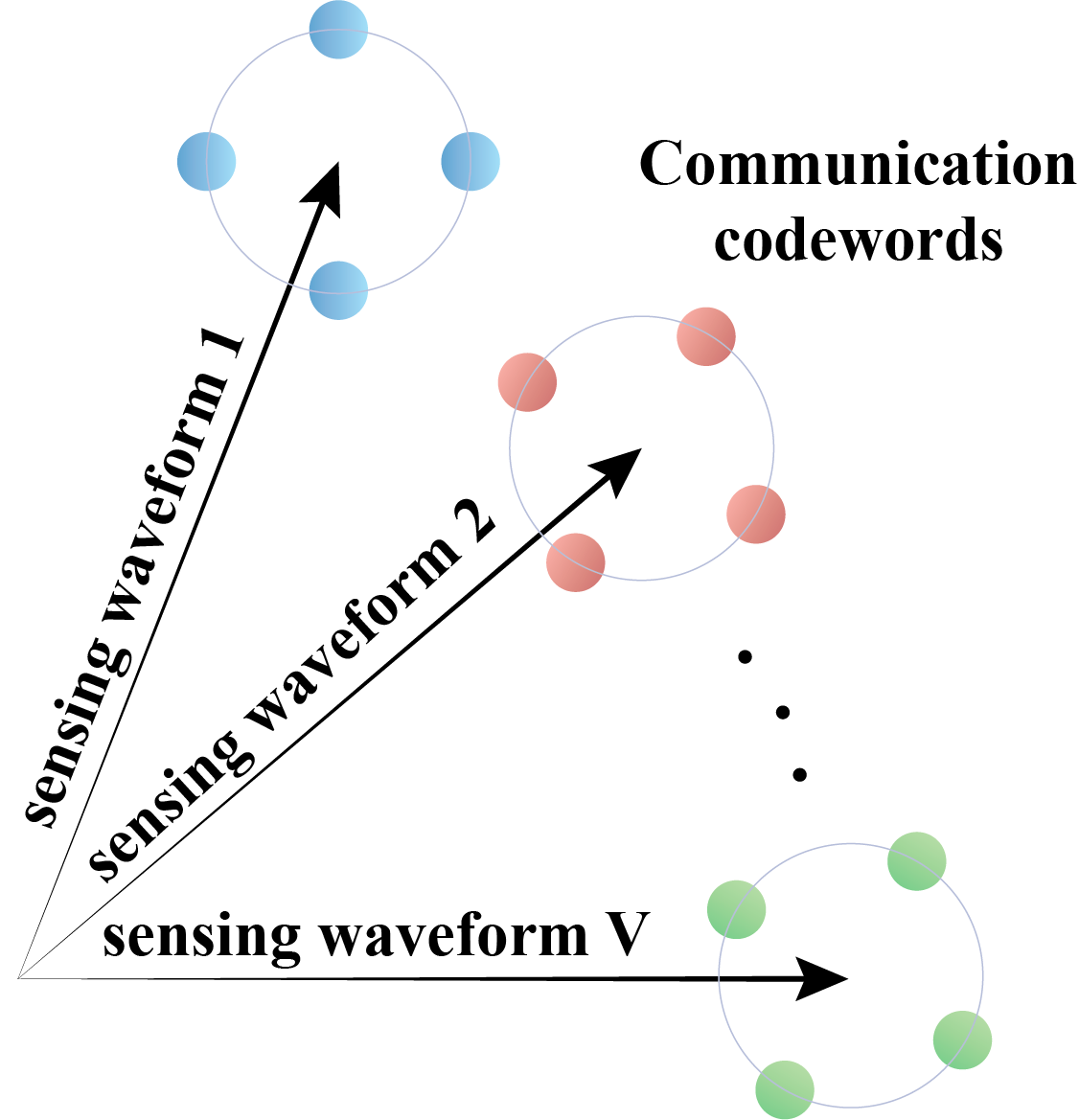}
  \caption{Waveform diversity and superposition.}\label{fig:diversity}
  \vspace{-0.5cm}
\end{figure}

\subsubsection{Spectrum Spreading Scheme}
Under this scheme, the communication message is modulated to a scalar symbol $x^{(c)}$ (e.g., a QAM). Then the dual-function signal is given by
\begin{eqnarray}
\vect x= x^{(c)} \vect x_{\text{opt}}^{(s)}.
\end{eqnarray}
Such a scheme is very similar to spread spectrum communications such as code-division-multiple-access (CDMA), in which the bandwidth of the communication signal is expanded by a spreading code (namely $\vect x_{\text{opt}}^{(s)}$ in the context of ISAC). 

\subsection{Waveform Unavailable at Receiver (WUR) Case}\label{sec:wur}
In this section, we consider the WUR case in which the sensing waveform $\vectsf x_{\text{opt}}^{(s)}$ is unknown to the communication receiver, but is confined to a set of possible waveforms  (similarly to a codebook in superposition coding). The major challenge is, when using the adaptive modulation approach introduced above, the communication receiver needs to first identify the selected sensing waveform $\vectsf x_{\text{opt}}^{(s)}$ (i.e., the cloud) and then to decode the communication message within the cloud. For simplicity, we assume that the communication channel matrix $\vectsf H_{c,\ell,m}$ is known to the communication receiver for each $m \in \{0, \ldots, M_c-1\}$ and $\ell \in \{0, \ldots, L-1\}$. 

As illustrated in Fig. \ref{fig:diversity}, for $V$ distinct sensing scenarios, the ISAC transceiver prepares a set of sensing waveforms $\{\vect x_{\text{opt},\ell,m,\nu}^{(s)}\}_{\nu=1}^{V}$ each with respect to corresponding power spectra $\left\{\P_{\ell,m,\nu}\right\}_{\nu=1}^{V}$ . Over the subcarrier $m$, for the received $L$ consecutive power spectra $\{Q_{\ell,m}\}_{\ell=0}^{L-1}$, the communication receiver needs to decode the communication message without the knowledge of $\{\vect x_{\text{opt},\ell,m,\nu}^{(s)}\}_{\nu=1}^{V}$ for each $\ell \in \{0, \ldots, L-1\}$.

To handle the unknown sensing waveform, we propose two methodologies corresponding to the three superposition coding schemes described in Section~\ref{sec:war}. The first one is the \emph{joint waveform estimation and decoding} methodology,  which we employ for the DPC and the spread spectrum schemes; and the second one is the \emph{cascaded waveform estimation and decoding} methodology, which we employ for the linear transform scheme. In the following we explain both methodologies in details.

\subsubsection{Joint Waveform Estimation and Decoding}
For the DPC and spread spectrum schemes the sensing waveform is only perturbed by the communication message. Therefore, we can carry out joint waveform estimation and decoding, and the final output is the decoded communication message. 
\begin{itemize}
\item DPC scheme: The estimates of the $L$ consecutive communication symbols $\vect w =\{\tilde w_{0},..., \tilde w_{L-1}\}$ and the waveform index $m$ are given by
\begin{IEEEeqnarray}{rCl}\label{eq:add_decode}
{(\hat m, \hat{\vect w})}=\arg\min_{m, \vect w}\sum_{l=0}^{L-1}\|\vect y_{\ell,m}^{(c)} -\vect{x}_{\ell,m}\|^2,
\end{IEEEeqnarray} 
where $\vect{x}_{\ell,m}$ is the transmitted DPC waveform determined by $\vect{x}_{\text{opt}, \ell,m}^{(s)}$ and communication message $\tilde w_{\ell}$ as explained in Section~\ref{sec:DPC}. 

\item Spread spectrum scheme: The decoding for this scheme follows  the decoding of CDMA receivers and is given by 
\begin{IEEEeqnarray}{rCl}\label{eq:scaling_decoding}
{(\hat m, \hat{\vect w})}=\arg\min_{m, \vect w}\sum_{l=0}^{L-1}\|\vect y_{\ell,m}^{(c)} -x_c(\tilde w_\ell)\vect{x}_{\text{opt},\ell}^{(s)}\|^2.
\end{IEEEeqnarray} 
\end{itemize}
When the number of possible sensing waveforms is not large, the above decoding procedure can be carried out by an exhaustive search over all possible sensing waveforms. 

\subsubsection{Cascaded Waveform Estimation and Decoding}
In the linear transform scheme (the OFDM scheme), the original sensing waveform is substantially distorted (although the power spectrum remains unchanged). Therefore, the sensing waveform $\vect{x}_{\text{opt}}^{(s)}$ needs to be estimated first. In this case, if a fixed PSK order is used, the communication receiver can directly demodulate the PSK symbols over the subcarriers after the DFT, regardless of the sensing waveform. However, when the PSK is adaptive, the sensing waveform (or equivalently the signal power spectrum) needs to be obtained, such that the PSK modulation order can be derived. Given that  the original sensing waveform is substantially distorted by the communication messages, we need to first identify the sensing waveform, and then demodulate the communication symbols according to the derived modulation order.

For each subcarrier $m$, one can estimate the modulation order from the corresponding received PSD normalized over all subcarriers, and then estimate the PSK modulation order $K$. 

\begin{itemize}
\item \emph{Nonparametric estimation}: When the distribution of noise is unknown, we simply calculate the average power over each subcarrier $m$:
\begin{IEEEeqnarray}{rCl}\label{eq:nonpara0}
\hat{\P}_m=\frac{1}{L}\sum_{\ell=0}^{L-1} Q_{\ell,m}-\hat{N}_0,
\end{IEEEeqnarray} 
where $Q_{\ell,m}$ is the measured power over the $m$-th subcarrier of the $\ell$-th symbol, and $\hat{N}_0$ is the estimated noise power. 

\item \emph{Maximum likelihood estimation}: A reasonable assumption is that the noise is circular symmetric Gaussian distributed with zero expectation and variance $\sigma_c^2$. Then, the noise power over each subcarrier is Rayleigh distributed. Therefore, the estimate of $\P_m$ can be obtained using maximum likelihood estimation:
\begin{IEEEeqnarray}{rCl}\label{eq:para0}
\hat \P_m&=&\arg\max_{\nu}\prod_{l=0}^{L-1}\prod_{m=0}^{M_c-1}|\P_{\ell,m,\nu}\text{Tr}(\vectsf H_{c,\ell,m}\vectsf H_{c,\ell,m}^H)-Q_{\ell,m}|\nonumber\\
&\times&e^{-\frac{|\P_{\ell,m,\nu}\text{Tr}(\vectsf H_{c,\ell,m}\vectsf H_{c,\ell,m}^H)-Q_{\ell,m}|^2}{2\sigma_c^2}}.
\end{IEEEeqnarray} 
\end{itemize}

\section{Sensing Waveform Optimization}\label{sec:design}
Based on the proposed broadcast channel of ISAC, in this section, we optimize the sensing waveform based on three objectives: Maximizing SCNR, minimizing the discrete-AF sidelobes under the effect of Doppler shift, and minimizing the expected ISL. 
\subsection{Maximizing SCNR}
In this section, for each $m$ and $\ell$,  we find the waveform that maximizes $\tilde \gamma_{\text{SCNR}, \ell,m} (t_0)$ in \eqref{eq:scnr2}.  For simplicity, we drop the subscript pair  $\ell,m$.   Our analysis follow the optimum waveform design argument in \cite[Chapter~6.3]{Pillai2011}. 

Let $\vectsf L_f(z)$ represents the minimum phase left-spectral factor matrix function associated with the total clutter and noise spectrum, i.e., $\vectsf S_{\ell,m} (w)$ in \eqref{eq:S}; then  
\begin{equation}
\vectsf S(w) = \vectsf L_f(e^{jw}) \vectsf L^*_f(e^{jw}) >0.
\end{equation}
The SCNR term in \eqref{eq:scnr2} thus can be simplified as 
\begin{IEEEeqnarray}{rCl}
\tilde \gamma_{\text{SCNR}}(t_0)
& :=&  \frac{\left | \frac{1}{2\pi} \int_{-\infty}^{\infty} \vectsf L_f(e^{jw}) \vect K^*(e^{jw}) dw\right |^2}{\frac{1}{2\pi} \int_{-\infty}^{\infty} \vectsf L_f(e^{jw}) \vectsf L_f^*(e^{jw})dw  }\\
&\stackrel{(i)}{\le} & \frac{1}{2\pi} \int_{-\infty}^{\infty}  \vect K^*(e^{jw}) \vect K(e^{jw}) dw  \\
&=& \int_{0}^{t_0} \vect \phi^*(t) \vect \phi(t) dt \label{eq:phi} \\
&:=& \tilde \gamma_{\text{SCNR},\max}, 
\end{IEEEeqnarray}
where $\vect \phi^*(t)$ represent the inverse transform of $\vectsf L_f^{-1}(e^{jw})\vectsf H_{t}(e^{jw}) \vect X^{(s)}(e^{jw})$, i.e., 
\begin{equation}
\vect \phi^*(t) \leftrightarrow \vectsf L_f^{-1}(e^{jw})\vectsf H_{t}(e^{jw}) \vect X^{(s)}(e^{jw})
\end{equation}
and 
\begin{equation}
\vect \phi^*(t_0-t)u(t) \leftrightarrow \vect K(e^{jw}).
\end{equation}
The inequality in $(i)$ follows by the Schwarz inequality  $|\int A(w)B(w) dw|^2 \le \int |A(w)|^2dw \int |B(w)|^2dw$ where equality is achieved  if and only if 
\begin{equation}
\vect K(z) \vectsf L_f^{-1} (z) = \textbf{I}_{N},
\end{equation}
where $\textbf I_{N}$ is the $N \times N$ identity matrix. 
Note that $\vect \phi(t)$ can be written as 
\begin{equation} \label{eq:psi2}
\vect \phi (t) = \int_{0}^{t_0} \vectsf \psi (t-\tau)\vect x^{(s)}(\tau) d\tau,
\end{equation}
where 
\begin{equation}
\vectsf \psi (t) \leftrightarrow \vectsf L_f^{-1}(e^{jw})\vectsf H_{t}(e^{jw}).
\end{equation}
Denote 
\begin{equation}\label{eq:omega2}
\vectsf \Omega(\tau_1, \tau_2) := \int_0^{t_0} \vectsf \psi^*(t-\tau_1) \vectsf \psi(t-\tau_1) dt.
\end{equation}
Combining \eqref{eq:psi2} and \eqref{eq:omega2} with \eqref{eq:phi} gives 
\begin{IEEEeqnarray}{rCl}
\lefteqn{\gamma_{\text{SCNR},\max} (t_0)} \notag \\
&=&  \int_0^{t_0} \vect \phi^*(t) \vect \phi(t) dt \\
&=& \int_{0}^{t_0} \int_{0}^{t_0} \vect x^{(s)^*} (\tau_1) \vectsf \Omega(\tau_1, \tau_2) \vect x^{(s)} (\tau_2) d\tau_1 d\tau_2\IEEEeqnarraynumspace \\
&=& \vect x^{(s)^*}_{t_0} \vectsf \Omega \vect x^{(s)}_{t_0}, \label{eq:omega}
\end{IEEEeqnarray}
where 
\begin{equation}
\vect x^{(s)}_{t_0} := \begin{bmatrix} \vect x^{(s)}(0) \\ \vect x^{(s)}(1) \\ \vdots \\ \vect x^{(s)}(t_0) \end{bmatrix}
\end{equation}
represents the column vector consisting of the transmit sensing signal set. 
Notice that $\vectsf \Omega$ in \eqref{eq:omega} can be written as  
\begin{equation}
\vectsf \Omega = \vectsf \phi^* \vectsf \phi,
\end{equation}
where 
\begin{equation}
\vectsf \phi := \begin{bmatrix} \vectsf \phi(0) & 0 & \ldots & 0 \\ \vectsf \phi(1) & \vectsf \phi(0) &\ldots & 0 \\ \vdots & \vdots& \ddots& \vdots \\ \vectsf \phi(t_0) & \vectsf \phi(t_0-1)& \ldots &\vectsf \phi(0) \end{bmatrix}.
\end{equation}

Equation \eqref{eq:omega} is used to design the normalized optimum sensing transmit signal set $\vect x^{(s)}_{\text{opt}}$. More specifically, the eigenvector $\vect x^{(s)}_{\text{opt}}$ associated with the largest eigenvalue of $\vectsf \Omega$ represents the normalized transmit sensing vector that maximizes the target echo while minimizing the clutter and noise components. In general, \eqref{eq:omega} represents a highly nonlinear problem due to the effect of clutter. In the absence of clutter, the above formulation can be solved exactly as in that case 
\begin{equation}
\vectsf S(w) = \vectsf G_n(w) = \vectsf L_n(e^{jw}) \vectsf L_n^*(e^{jw}),
\end{equation}
where $\vectsf L_n(z)$ represents the minimum phase factor associated with the noise spectral density matrix $\vectsf G_n(w)$. As a result, the optimum transmit sensing vector is given by the eigenvector associated with the largest eigenvalue of the known positive definite matrix $\vectsf \Omega(\tau_1,\tau_2)$ defined in \eqref{eq:omega2} but with $\vectsf \psi(t) \leftrightarrow \vectsf L_n^{-1}(e^{jw})\vectsf H_t(e^{jw})$. In the presence of clutter, the optimum sensing transmit signal $\vect x^{(s)}_{\text{opt}}$ can be obtained by solving \eqref{eq:omega} in an iterative manner. To this end, let any causal vector $\vect x^{(s)}_0$ with each element of duration $t_0$ and the total energy $E_s$ be the initialization vector. At stage $k$, assume $\vect x^{(s)}_k \leftrightarrow \vect X^{(s)}_k$  is the solution.  Let $\vectsf L_{k}(e^{jw})$ be the minimum phase factor associated with the equation 
\begin{equation}\label{eq:Lk}
\vectsf L_{k}(e^{jw}) \vectsf L_{k}^*(e^{jw}) = \sum_{g = 1}^{\mathcal G} \vectsf H_{g}(w) \vect X_k^{(s)}(e^{jw})\vect X_k^{(s)^*}(e^{jw}) +  \vectsf G_n(w)
\end{equation}
and define 
\begin{equation}
\vectsf \psi_k(t) \leftrightarrow \vectsf L_{k}^{-1}(e^{jw})\vectsf H_t(e^{jw}).
\end{equation}
Compute 
\begin{equation}
\vectsf \phi_k := \begin{bmatrix} \vectsf \phi_k(0) & 0 & \ldots & 0 \\ \vectsf \phi_k(1) & \vectsf \phi_k(0) &\ldots & 0 \\ \vdots & \vdots& \ddots& \vdots \\ \vectsf \phi(t_0) & \vectsf \phi_k(t_0-1)& \ldots &\vectsf \phi_k(0) \end{bmatrix},
\end{equation}
and define 
\begin{equation}
\vectsf \Omega_k = \vectsf \phi^*_k \vectsf \phi_k > 0.
\end{equation}
Hence, the normalized eigenvector $\vect u_1^{(k)}$ associated with the largest eigenvalue $\lambda_1^{(k)}$ of the equation 
\begin{equation}
\vectsf \Omega_k \vect u_1^{(k)} = \lambda_1^{(k)} \vect u_1^{(k)}
\end{equation}
coincides with the optimum solution at stage $k$. Ideally, we should have the identity 
\begin{equation}
\vect x^{(s)}_k = \sqrt{E_s} \vect u_1^{(k)}.
\end{equation}
However, that may not always be the case and the difference, given by 
\begin{equation}
\vect \epsilon_k = \vect x^{(s)}_k - \sqrt{E_s} \vect u_1^{(k)}
\end{equation}
can be used to update $ \vect x^{(s)}_k$ by suitably combining it with $\vect u_1^{(k)}$. Towards this, we define the update rule as
\begin{equation}
\vect x^{(s)}_{k+1} = \frac{\vect x^{(s)}_k + ||\vect \epsilon_k|| \vect u_1^{(k)}}{\left \|\vect x^{(s)}_k + ||\vect \epsilon_k|| \vect u_1^{(k)}\right \|}.
\end{equation}
We then use $\vect x^{(s)}_{k+1}$ to update $\vectsf L_{k+1}(z)$ from $\vectsf L_{k}(z)$ in \eqref{eq:Lk}. The entire algorithm should be repeated until the error $\| \vect \epsilon_k\|$ is acceptably small. The optimum transmit sensing signal vector thus is
\begin{equation}
\vect x^{(s)}_{\text{opt}} = \lim_{k \to \infty} \vect x^{(s)}_k \label{eq:57}
\end{equation}
and the maximum SCNR is given by 
\begin{equation}
\tilde \gamma_{\text{SCNR},\max} = \lim_{k \to \infty} \lambda_1(k).
\end{equation}

\subsection{Minimizing the Discrete-AF Sidelobes} 
In this section, the objective is to design the sensing signal sequence $\{\tilde x_{\ell,m}^{(s)}\}$, $m \in \{0, \ldots, M_c-1\}$, $\ell \in \{0, \ldots, L-1\}$, so as to minimize the sidelobes of the discrete AF in \eqref{eq:AFd}. Specifically, the objective is 
\begin{equation}
\min_{\{\tilde x_{\ell,m}^{(s)}\}} \;  \sum_{d \in \mathcal A} \sum_{\nu \in \mathcal B} | \chi(d,\nu) |^2,
\end{equation}  
where $\tilde x_{\ell,m}^{(s)} := \vect a_t^H(\theta_0) \vect x_{\ell,m}^{(s)}$. 
Assume that the time delay and the Doppler frequency sets of interests are given by 
\begin{IEEEeqnarray}{rCl}
\mathcal A &:=& \{0, \pm 1, \ldots, \pm A\}, \label{eq:setA} \\
\mathcal B &:=& \{0, \pm 1, \ldots, \pm B\}. \label{eq:setB}
\end{IEEEeqnarray}
For each $m \in \{0, \ldots, M_c-1\}$, define a set of $A$ sequences
\begin{IEEEeqnarray}{rCl} \label{eq:set}
\{x_{m, a}(\ell) = \tilde x_{\ell,m}^{(s)}e^{j2\pi (a-1)\frac{\ell}{L}}\}, \end{IEEEeqnarray}
with $\ell \in \{0, \ldots, L-1\}$ and $a \in \{1, \ldots, A\}$. Note that $\{x_{m, a}(\ell)\}$ are zero when $\ell \notin \{0, \ldots, L-1\}$. Let $\{r_{a_1a_2}(d)\}$ denote the correlation between $\{x_{m, a_1}(\ell)\}$ and $\{x_{m, a_2}(\ell)\}$ and is given by
\begin{IEEEeqnarray}{rCl}
r_{a_1,a_2}(d) &=& \sum_{\ell = 0}^{L-1} x_{m,a_1}(\ell) x_{m,a_2}^*(\ell-d) \\
& = & e^{\frac{j2\pi (a_1-1)d}{L}} \sum_{\ell = 0}^{L-1} \tilde x_{\ell,m}^{(s)} ( \tilde x_{\ell,m-d}^{(s)} )^* e^{\frac{j2\pi(\ell -d)(a_2-a_1)}{L}} \IEEEeqnarraynumspace
\end{IEEEeqnarray}
with $a_1, a_2 \in \{1, \ldots, A\}$. It is straightforward to show that all values of $|\chi (d,\nu)|$ ($d \in \mathcal B$, $\nu \in \mathcal A$), are contained in the set $\{ |r_{a_1,a_2}(d) |\}$ with $a_1, a_2  \in \{1, \ldots, A\} $, $d \in \mathcal B$ and  $m \in \{0, \ldots, M_c-1\}$.  Therefore, by minimizing the correlations of the sequence set in \eqref{eq:set}, we can equivalently minimize the discrete-AF sidelobes. Define
\begin{equation}
\vectsf X = [\vectsf X_1 \ldots \vectsf X_A] \in \mathbb C^{(L+B-1)M_c \times BA} 
\end{equation}
where 
\begin{equation}
\vectsf X_a := \begin{bmatrix} \vect X_{M_c,a}(0)&& 0 \\
\vdots& \ddots & \\ 
\vdots& & \vect X_{M_c,a}(0) \\
\vect X_{M_c,a}(L-1)&& \vdots \\
& \ddots & \vdots \\
0& & \vect X_{M_c,a}(L-1)
 \end{bmatrix}_{(L+B-1)M_c \times B}
\end{equation}
with $\vect X_{M_c,a}(\ell):= [x_{0,a}(\ell) \; x_{1,a}(\ell)\;  \ldots \; x_{M_c-1,a}(\ell) ]^T$. It is easy to show that all $\{r_{a_1a_2}(d)\}$  appear in the matrix $\vectsf X^H \vectsf X$. Also, note that the diagonal elements of $\vectsf X^H \vectsf X$ are equal to $E_s:= \sum_{m = 0}^{M_c-1}\sum_{ \ell = 0}^{L-1} |\tilde x_{\ell,m}|^2$. Therefore, the correlations of the sequence set in \eqref{eq:set} can be made small by minimizing the following quantity: 
\begin{equation}
Q := || \vectsf X^H \vectsf X - E_s\vect I_{AB}||.
\end{equation}
Note that, $Q$ equals zero if the matrix $\vect X$ is a semi-unitary matrix scaled at $\sqrt{E_s}$. Hence, minimizing $Q$ can be simplified to 
\begin{subequations} \label{eq:69}
\begin{IEEEeqnarray}{rCl}
\min_{\vectsf X, \vectsf U} \; && || \vectsf X - \sqrt{E_s} \vectsf U||^2, \\
\text{s.t.:} \: && \vectsf U^H \vectsf U = \vect I.
\end{IEEEeqnarray}
\end{subequations}
To solve this optimization problem, we employ the multi-cyclic algorithm in \cite{Stoica2008, He2012}. To this end, we assume that the sensing signals are unit-modulus and  perform the following steps:
\begin{itemize}
\item Randomly initialize the sequence $\{\tilde x^{(s)}_{\ell,m}\}$. 
\item For fixed $\vectsf X$, calculate the minimizer $\vectsf U$, which is given by \cite{He2012} 
\begin{equation}
\vectsf U = \vectsf U_2 \vectsf U_1^H
\end{equation}
where matrices $\vectsf U_1 \in \mathbb C^{BA \times BA}$ and $\vectsf U_2 \in \mathbb C^{(L+B-1)M_c\times BA}$ are the results of single value decomposition (SVD) of $\vectsf X^H$, i.e., $\vectsf X^H = \vectsf U_1 \vectsf \Sigma \vectsf U_2^H$. 
\item For fixed $\vectsf U$, the objective in \eqref{eq:69} can be written as 
\begin{IEEEeqnarray}{rCl}
\lefteqn{||\vectsf X - \sqrt{E_s} \vectsf U||^2 } \notag \\&=& \sum_{m = 0}^{M_c-1} \sum_{\ell = 0}^{L-1} \sum_{j = 1}^{AB} |\mu_{\ell,m, j} \tilde x_{\ell,m}^{(s)} - f_{m, \ell, j}|^2\IEEEeqnarraynumspace\\ 
& = & \tilde c - 2\sum_{m = 0}^{M_c-1} \sum_{\ell = 0}^{L-1} \text{Re} \left (\sum_{j = 1}^{AB} \mu^*_{\ell,m, j}f_{m, \ell, j}\right) (\tilde x_{\ell,m}^{(s)})^*
\end{IEEEeqnarray}
where $\tilde c$ is a constant that is independent from $\{\tilde x_{\ell,m}^{(s)}\}$, $\{f_{m, \ell, j}\}$ are the corresponding elements of $\sqrt{E_s} \vectsf U$, and for each $a \in \{1, \ldots, A\}$, $\{\mu_{\ell,m, j}\}$ are given by
\begin{equation}
[\mu_{\ell,m,(a-1)B+1} \ldots \mu_{\ell,m, aB}] = e^{j2\pi (a-1)\frac{\ell}{L}}\vec{\vectsf 1}_B, 
\end{equation} 
where $\vec{\vectsf 1}_B$ is an all-one vector of length $B$. 
\item For each $m \in \{0, \ldots, M_c-1\}$ and $\ell \in \{0, \ldots, L-1\}$, the minimizer $\tilde x_{\ell,m}^{(s)}$ then is obtained by 
\begin{equation}\label{eq:65n}
\tilde x_{\ell,m}^{(s)} = e^{j \phi_{\ell,m}}
\end{equation}
where 
\begin{equation}
\phi_{\ell,m} = \arg \left ( \sum_{j = 1}^{AB} \mu^*_{\ell,m, j}f_{m, \ell, j} \right). 
\end{equation}
\end{itemize}
The above steps are repeated until convergence. The optimum sensing waveform is then given by 
\begin{equation}
\vect x^{(s)}_{\text{opt},\ell,m} = \vect a_t(\theta_0) \tilde x_{\ell,m}^{(s)},
\end{equation} 
with $\tilde x_{\ell,m}^{(s)}$ as in \eqref{eq:65n}.

\begin{algorithm}[t!]
	\caption{Sensing waveform synthesis in CC}\label{alg:water}
	\begin{algorithmic}[1]
		\STATE{Given the average subcarrier power allocation of communication signals $\{P_{\ell,m}^{(c)}\}$ for $m\in \{0,...,M_c-1\}$ and $\ell \in \{0, \ldots, L-1\}$.}
		\STATE{Set the initial value of the Lagrange multiplier $\lambda$, the threshold $\gamma$ and the step $\epsilon$.}
		\WHILE{$\tilde{\P}_s<\P_s-\gamma$}
   		        \STATE{Set $\P_{\ell,m}^{(s)}=(\lambda-\P_{\ell,m}^{(c)})^+$, for $m \in \{0,...,M_c-1\}$ and $\ell \in \{0, \ldots, L-1\}$.}
		        \STATE{Calculate $\tilde{\P}_s=\sum_{m=0}^{M_c-1}\sum_{\ell = 0}^{L-1} \P_{\ell,m}^{(s)}$.}
		        \STATE{Set $\lambda=\lambda+\epsilon$.}
		\ENDWHILE
		\STATE{For $m\in \{0,...,M_c-1\}$ and $\ell \in \{0, \ldots, L-1\}$, set the sensing signals $\vect x_{\ell,m}^{(s)}$ according to the power $\P_{\ell,m}^{(s)}$ with random phases.}
	\end{algorithmic}
\end{algorithm}

\subsection{Minimizing the Expected ISL} \label{sec:ISL}
In this section, the objective is to design the sensing waveform so as to minimize   the expectation of the ISL (over the randomness of the communication signal $\vect x^{(c)}$) and the constraint is the power allocated to the sensing signal, i.e., 
\begin{IEEEeqnarray}{rCl}\label{eq:CFP}
\min_{\vect x^{(s)} } &&\;\;  \mathbb E\left [\xi_{\text{ISL}}(\vect x) \right ]\nonumber\\
\text{s.t.} &&\;\;  \sum_{m = 0}^{M_c -1}\sum_{\ell = 0}^{L-1} \mathbb E\left [\| \vect x_{\ell,m}^{(s)}\|^2 \right ]\leq \P_s.
\end{IEEEeqnarray}

The optimization problem can be solved using numerical approaches (e.g., gradient descent similar to \cite{He2012}), since an explicit solution is prohibitive. In this paper, we propose a novel  approach to solve this optimization problem for the case with no Doppler-shift. Specifically, we consider the AF only along the $\tau$-axis, i.e., the range. Equivalently, we can express the AF as  the following  autocorrelation function $r$:
\begin{eqnarray}
r(\tau)=\int_{\tau}^{T_p}x_0(t)x_0^*(t-\tau)dt,
\end{eqnarray}
where $T_p := \frac{T_c}{M_c}$. 

 We first notice that the Fourier transform of the autocorrelation function $r(\tau)$ corresponds to the power spectral density (PSD) $P_X(\omega)$ of the signal $x_0$. Then, Parseval's identity states that
\begin{eqnarray}\label{eq: Parseval}
\int_0^\infty r^2(\tau) = \frac{1}{4\pi} \int_{\omega_c+\frac{\mathsf B}{2}}^{\omega_c-\frac{\mathsf B}{2}} P_X^2(\omega)d\omega,
\end{eqnarray}
where $w_c:= 2\pi f_c$ with $f_c$ being the carrier frequency. The left-hand side (LHS) of \eqref{eq: Parseval} can be approximated by
\begin{eqnarray}\label{eq:LHS}
\int_0^\infty r^2(\tau)\approx T_c \left (\xi_{\text{ISL}}(\vect x) +r(0)\right)=T_c \left (\xi_{\text{ISL}}(\vect x) +\P_t\right).
\end{eqnarray}
The right-hand side (RHS) of \eqref{eq: Parseval} is given by
\begin{eqnarray}\label{eq:RHS}
\int_{\omega_c+\frac{\mathsf B}{2}}^{\omega_c-\frac{ \mathsf B}{2}} P_X^2(\omega)dw=\mathsf B \text{Var}(P_X(\omega_c))+\frac{\P_t^2}{\mathsf B}.
\end{eqnarray}

Therefore, despite the approximation, it is reasonable to minimize the variance of the PSD, in order to minimize the ISL. Since the sensing waveform $\vect x^{(s)}$ is synthesized before the formation of the communication signal $\vect x^{(c)}$, $\vect x^{(s)}$ is designed adaptively to the statistics of $\vect x^{(c)}$. Next, we propose a simple water-filling approach summarized in Algorithm \ref{alg:water} to solve this optimization problem. The underlying rationale is that water-filling can effectively reduce the variance of the PSD, thereby minimizing the ISL, as disclosed in (\ref{eq:LHS}) and (\ref{eq:RHS}). The output of the algorithm is the optimum sensing waveform $\vect x^{(s)}_{\text{opt}}$. The optimality of the water-filling scheme for minimizing the PSD variance is established in the following Proposition~\ref{prop:optimality} and the proof is omitted in this paper.
\begin{proposition}\label{prop:optimality}
Given the sensing power constraint $\P_s$, the water-filling scheme in Algorithm \ref{alg:water} minimizes the PSD variance, and thus the ISL.
\end{proposition}

\section{Numerical Analysis}\label{sec:numerical}

In this section, we present the numerical results for our proposed superposition coding schemes for ISAC.
\subsection{WAR and WUR Decoding Comparison}
\begin{figure}[t!]
  \centering
  \begin{subfigure}{0.49\linewidth }
  \centering
  \includegraphics[width=1\textwidth]{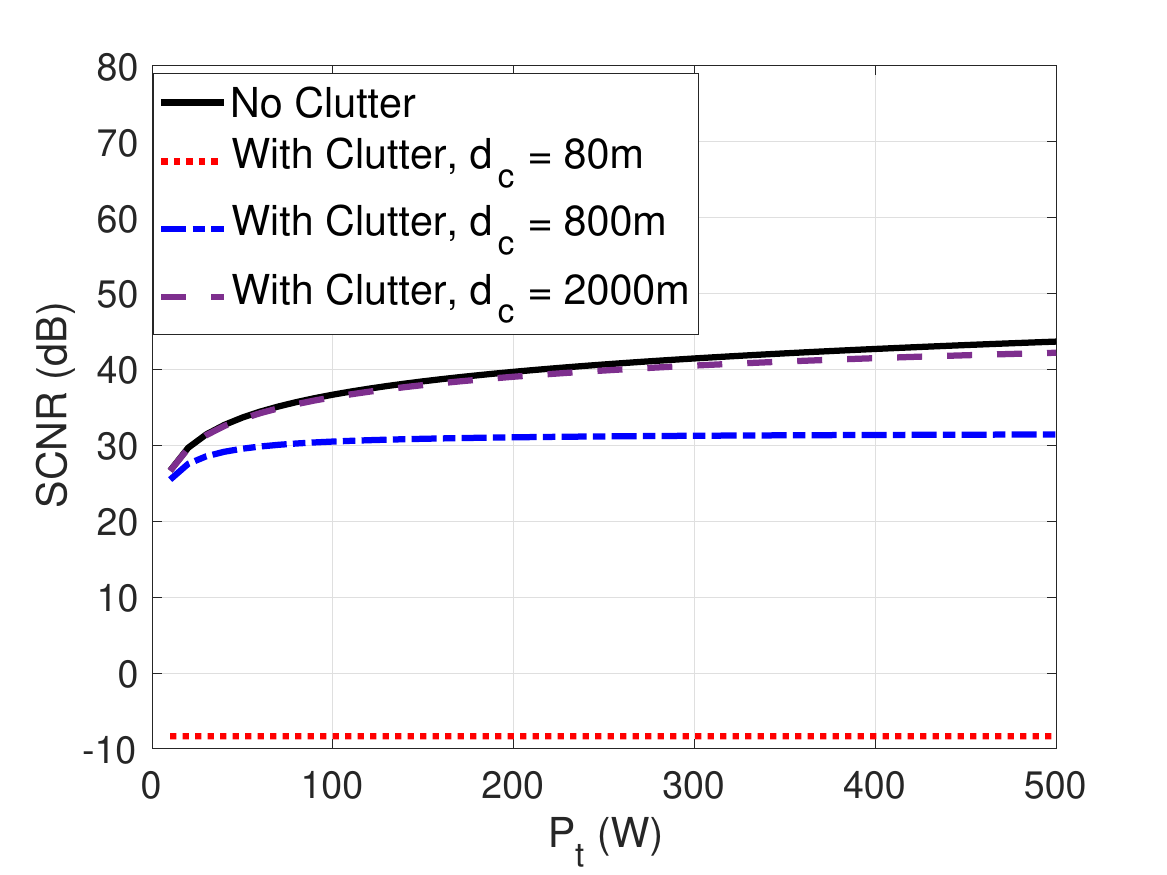}
  \caption{$d = 100$m, $r = 0.3$m}
  \label{fig:scnra}
  \end{subfigure}
   \begin{subfigure}{0.49\linewidth }
   \centering
  \includegraphics[width=1\textwidth]{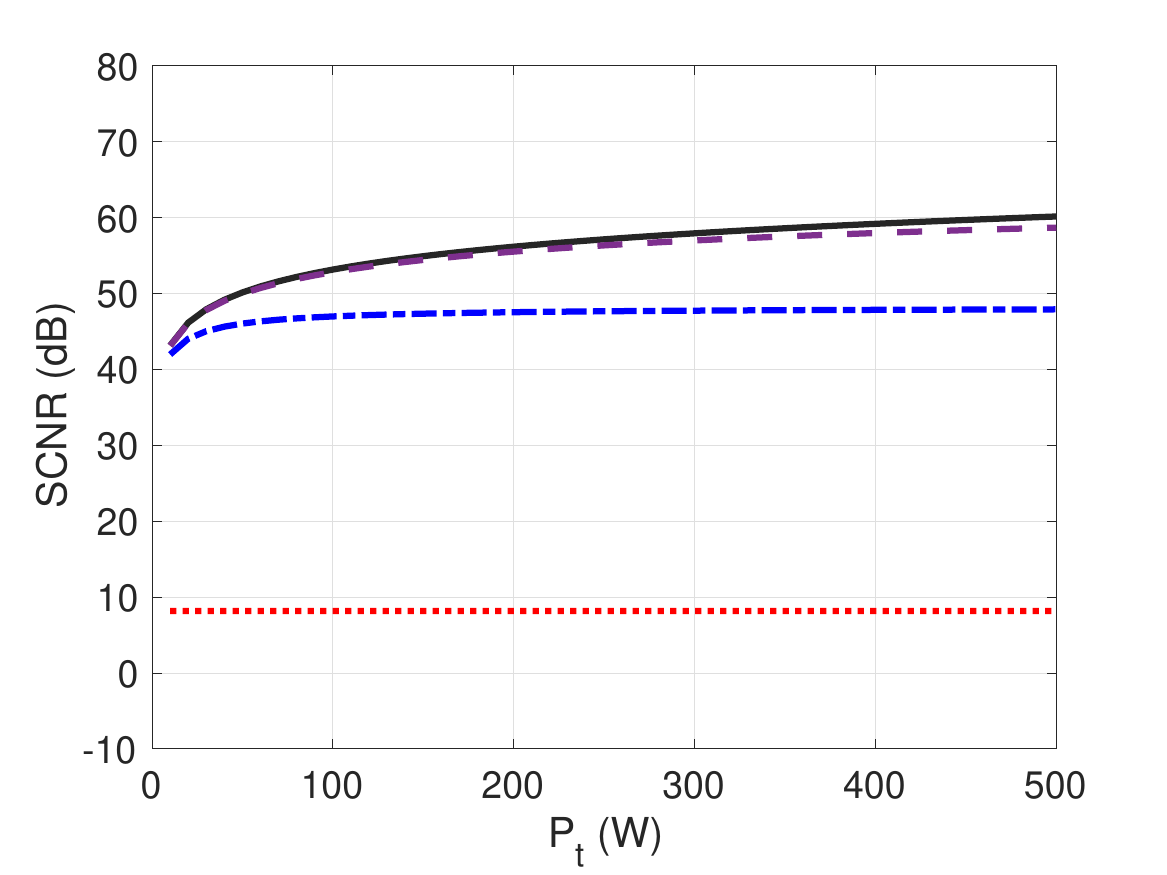}
  \caption{$d = 100$m, $r = 2$m}
  \label{fig:scnrb}
  \end{subfigure}
  
   \begin{subfigure}{0.49\linewidth }
   \centering
  \includegraphics[width=1\textwidth]{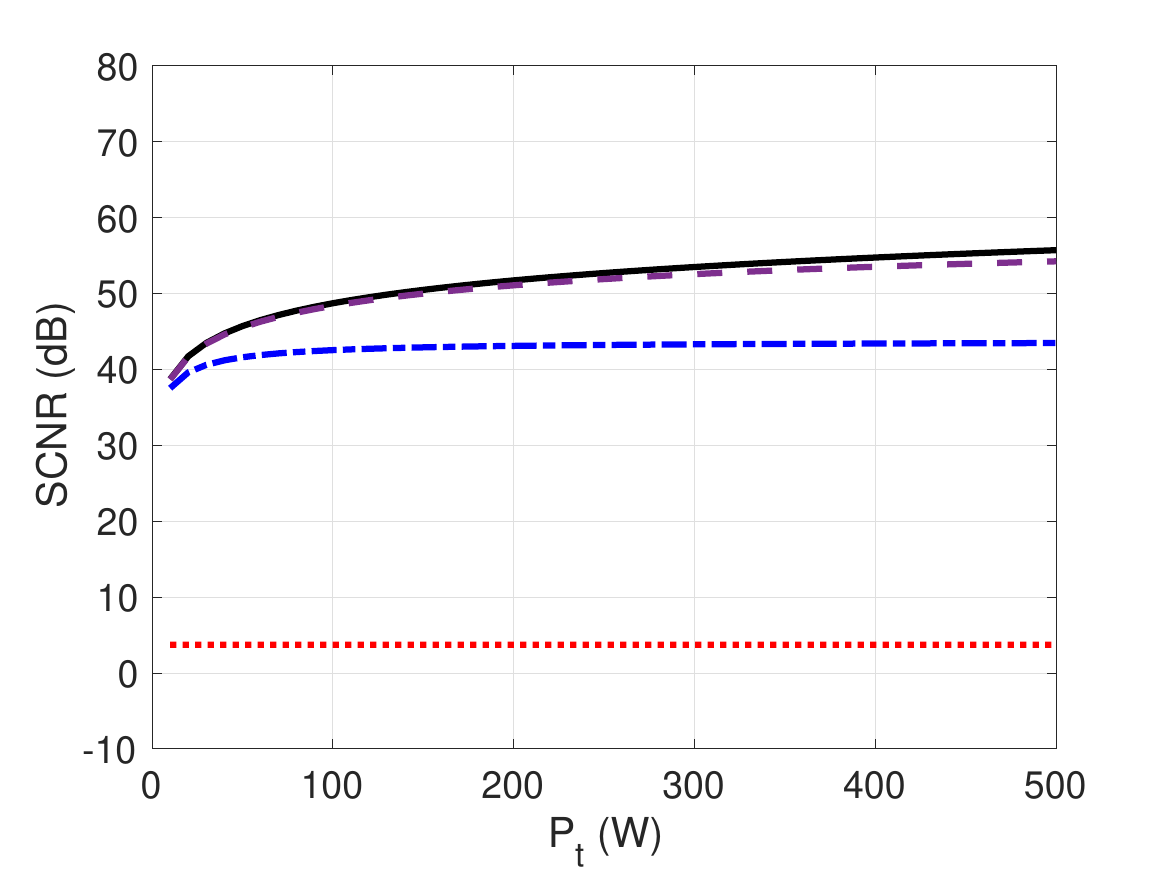}
  \caption{$d = 50$m, $r = 0.3$m}
  \label{fig:scnrc}
  \end{subfigure}
   \begin{subfigure}{0.49\linewidth }
   \centering
  \includegraphics[width=1\textwidth]{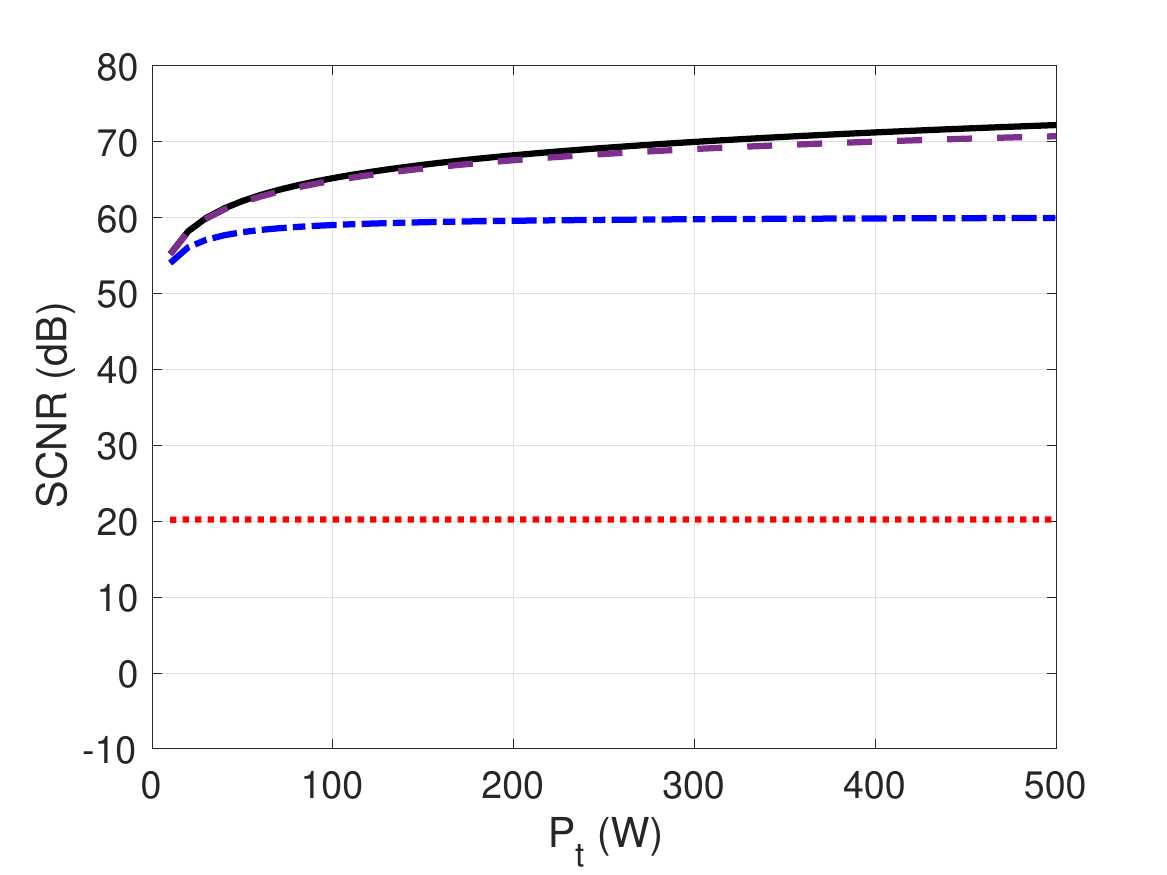}
  \caption{$d = 50$m, $r = 2$m}
  \label{fig:scnrd}
   \end{subfigure}
  \caption{Maximum SCNR vs. transmit power $P_t$ for various targets with different clutter configurations for $G_t = G_r = 30$.}
  \label{fig:scnr}
  \vspace{-0.3cm}
\end{figure}

We consider a 2-by-2 MIMO setup at the 2.5GHz band and a bandwidth of 400MHz. We first consider the maximum SCNR as the design metric for the optimum sensing waveform. We consider the target to be a sphere with radius $r  \in \{0.3m, 2m\}$  located at a distance of $d \in \{50m, 100m\}$ from the ISAC receiver. We model the target response using the sinc-sphere  approximation of Mie scattering, which captures the key frequency‑selective behavior of a uniform sphere without summing the full infinite series. Specifically, we assume that the RCS of the target is given by
\begin{equation}
\delta_{\text{RCS}} = \pi r^2 P_t G_t G_r\lambda^2 F^2(x),
\end{equation}
where $G_t$ and $G_r$  are the transmit and receive antenna gains, $\lambda$ is the wavelength, and $F(x)$ is the sinc-sphere form factor given by 
\begin{equation}
F(x) = \frac{3\left (\sin(x) - x\cos(x)\right)}{x^3}
\end{equation}
with $x = \frac{2\pi r f_c}{c}$.
We also consider a clutterer in a shape of a sphere with radius $r_c$ that is located at a distance of $d_c$ of the ISAC receiver. For the four combinations of the target distance and its radius, we generate four sensing waveforms, using \eqref{eq:57}. In Fig.~\ref{fig:scnr}, we plot the maximum SCNR as a function of the transmit power $\P_t$ for different targets. To model the clutter we assume $r_c = 0.5m$ and change the clutter distance $d_c$ from $80m$ to $2000m$. We consider a flat noise PSD of $-174$dBm/Hz. As can be seen from this figure, the maximum SCNR increases when the target is large or is located closer to the ISAC receiver. We also observe that a clutterer that is located close to the receiver can significantly decrease the SCNR of the target at the receiver. 


\begin{figure}[t!]
  \centering
  \begin{subfigure}{0.49\linewidth }
  \centering
  \includegraphics[width=1\textwidth]{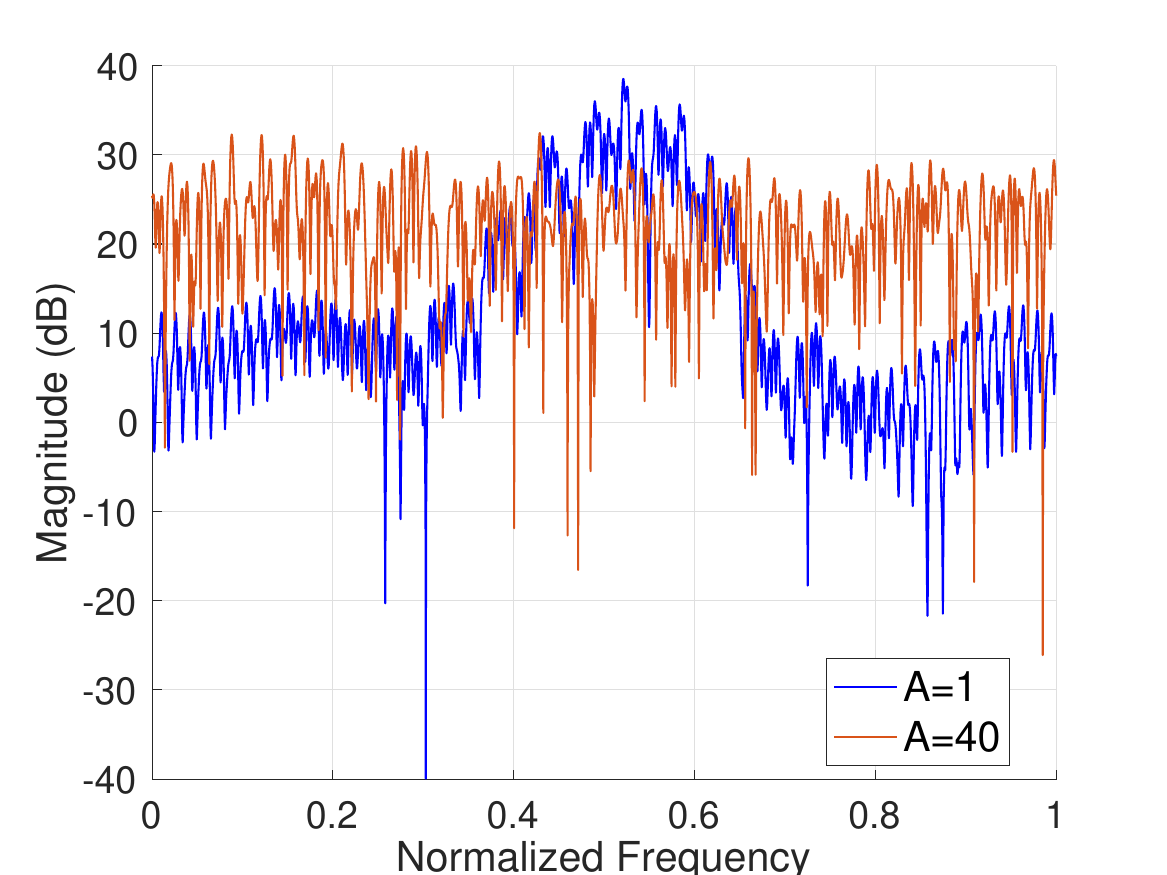}
  \caption{}
  \label{fig:AFa}
  \end{subfigure}
   \begin{subfigure}{0.49\linewidth }
   \centering
  \includegraphics[width=1\textwidth]{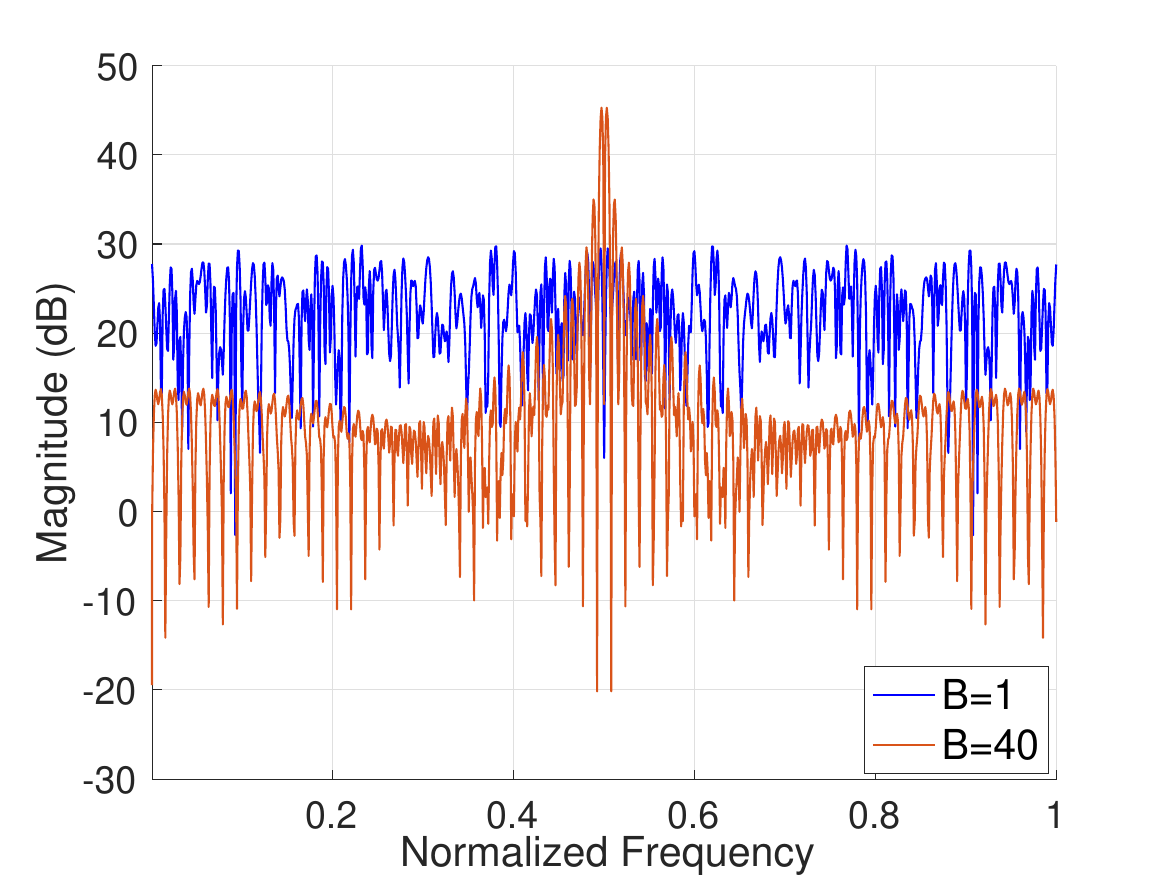}
  \caption{}
  \label{fig:AFb}
  \end{subfigure}
  
   \begin{subfigure}{0.49\linewidth }
   \centering
  \includegraphics[width=1\textwidth]{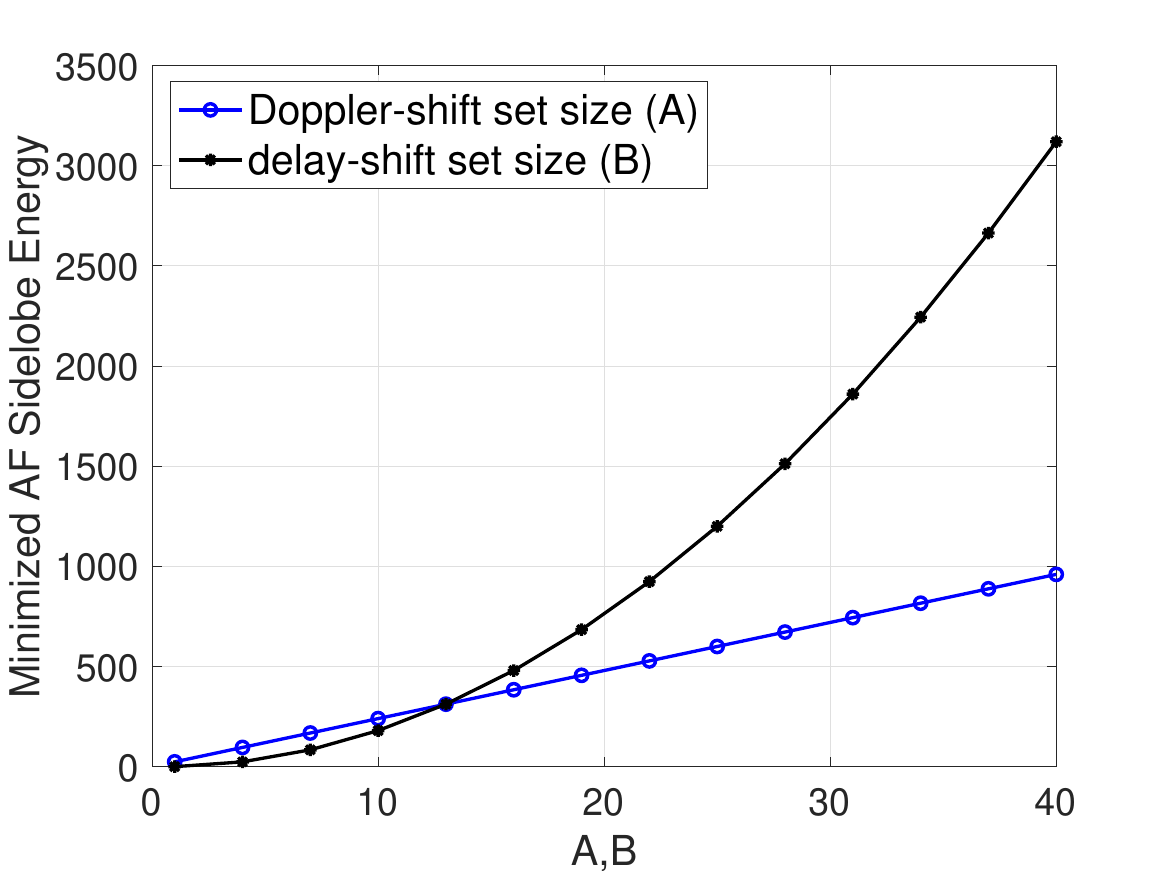}
  \caption{}
  \label{fig:AFc}
  \end{subfigure}
   \begin{subfigure}{0.49\linewidth }
   \centering
  \includegraphics[width=1\textwidth]{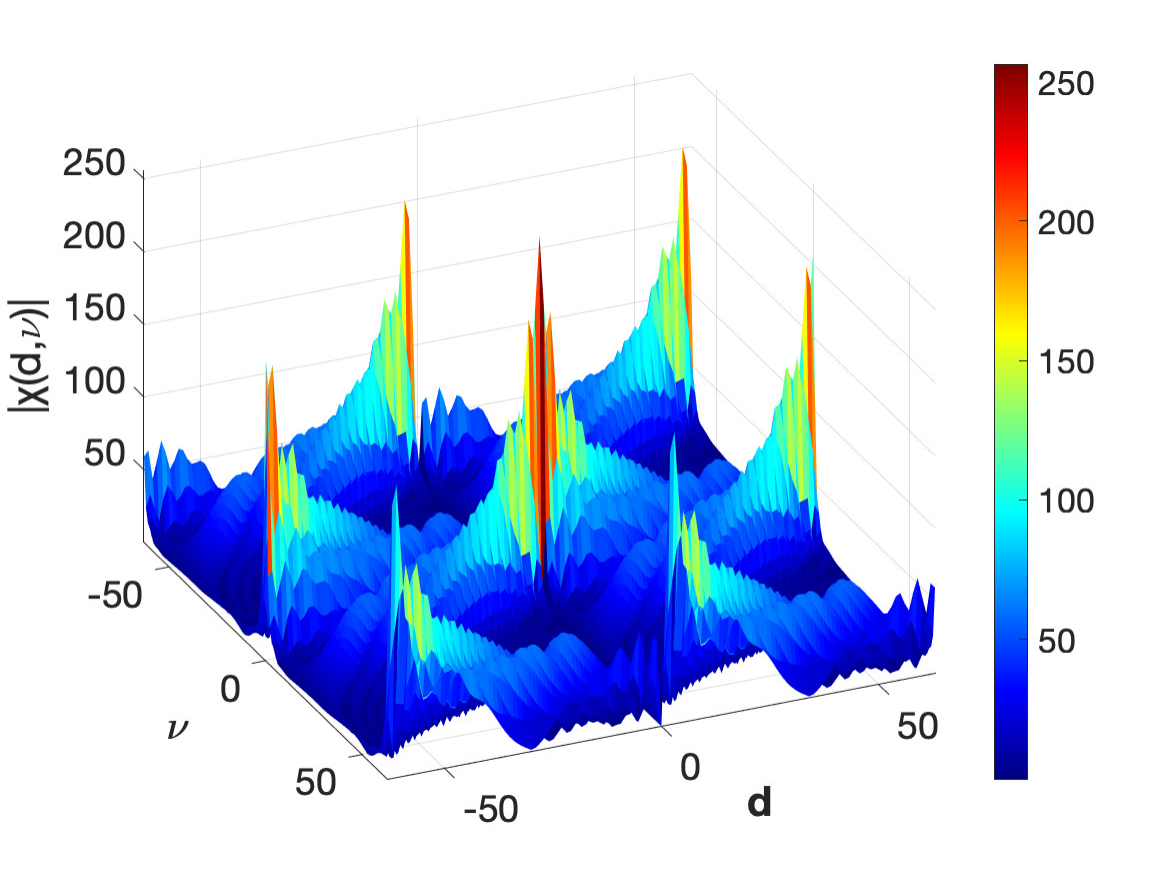}
  \caption{}
  \label{fig:AFd}
   \end{subfigure}
   
  \caption{Optimum sensing waveform for minimizing discrete-AF sidelobes: a) waveform spectra for $B = 4$ and $A = \{1,40\}$,  b) waveform spectra for $A = 1$ and $B = \{1,40\}$, c) minimized AF sidelobe energy for different values of $A$ and $B$, d) $|\chi(d, \nu)|$ vs. range and Doppler bins $d$ and $\nu$ for $A =1$ and $B = 4$.}
  \label{fig:AF}
  \vspace{-0.3cm}
\end{figure}

\begin{figure*}[t!]
  \centering
  \begin{subfigure}{0.32 \linewidth}
  \centering
  \includegraphics[width=0.95\textwidth]{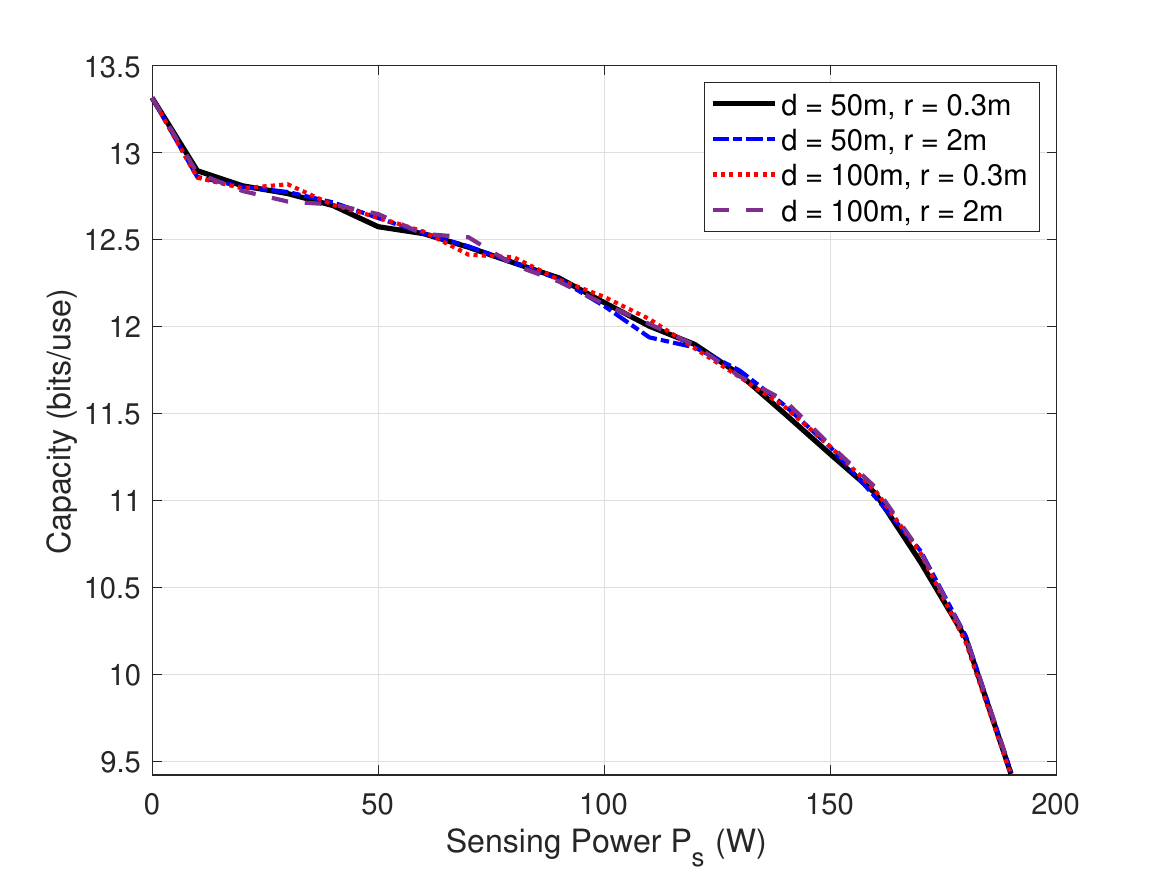}
  \caption{Capacity  vs   $\P_s$ under the DPC scheme
  .}\label{fig:dpc}  \end{subfigure}
  \begin{subfigure}{0.32 \linewidth}
    \centering
  \includegraphics[width=0.95\textwidth]{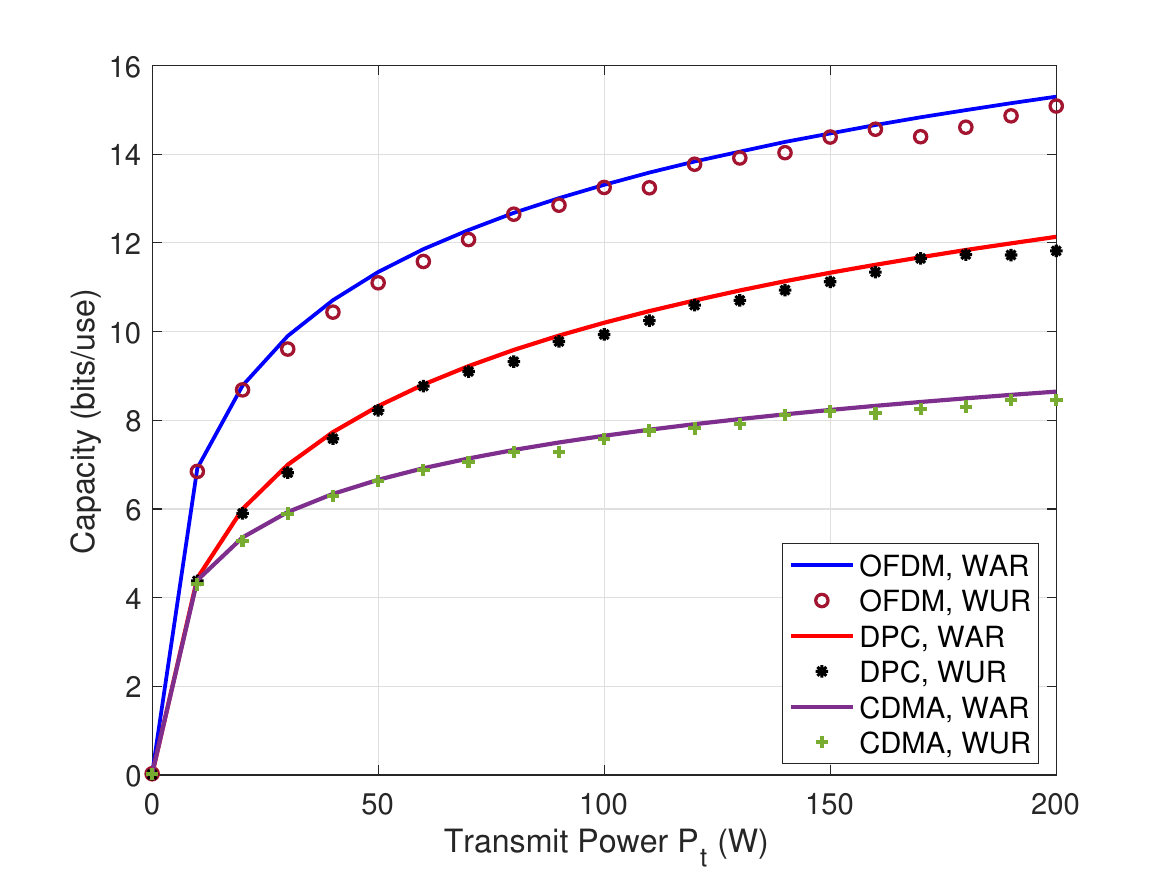}
  \caption{Capacity vs $\P_t$ for OFDM, DPC, CDMA.}
   \label{fig:ofdmdpc}
  \end{subfigure}
\begin{subfigure}{0.32 \linewidth}
  \centering
  \includegraphics[width=0.95\textwidth]{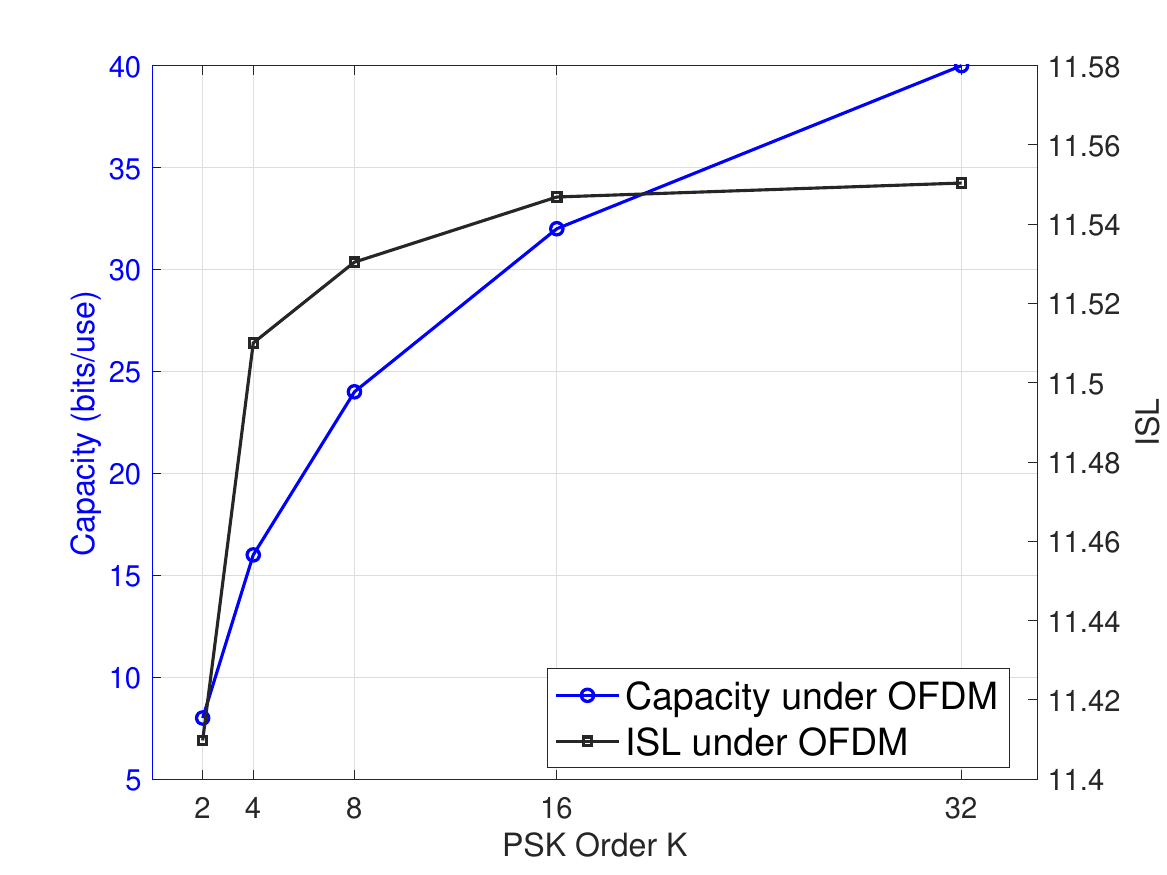}
  \caption{Capacity vs ISL for various $K$ and OFDM}
  \label{fig:ofdmISL}
  \end{subfigure}
\caption{Performance analysis of different superposition schemes: a) Capacity  vs. the sensing transmit power $\P_s$ for various target scenarios under the DPC scheme, b) Capacity vs.  total transmit power $\P_t$ under the OFDM, DPC and CDMA schemes for WAR and WUR scenarios, c) Capacity vs. ISL for different modulation order $K$ under the OFDM scheme.}
  \label{fig:ofdmdpc1}
\end{figure*}

In Fig.~\ref{fig:AF}, we consider minimizing the discrete-AF sidelobes as the design metric for the optimum sensing waveform. For this figure, we set $L = 64$, $M_c = 4$ and the convergence criterion of the algorithm is set at $10^{-6}$. In the employed multi-cyclic algorithm, the sequence $\{\tilde x^{(s)}_{\ell,m}\}$ is initialized using a Golomb sequence of length $2^n-1$ with $n = 9$. The Doppler frequency and the time delay sets $\mathcal A$ and $\mathcal B$ are defined in \eqref{eq:setA} and \eqref{eq:setB}, respectively.  Note that $|\mathcal A| = 2A + 1 $ and $|\mathcal B| = 2B +1$ where $A$ and $B$ are positive integers. Fig.~\ref{fig:AFa} shows the optimum sensing waveform spectra for the case where $B = 4$ and  $A \in \{1, 40\}$. As can be seen from this figure,  with an increase in the number of Doppler‐shift bins $A$, the optimum waveform shows increased spectral ripples and flatter envelopes in the frequency domain. This is due to the fact that in this case, energy is distributed  more evenly across the entire band to null out multiple Doppler sidelobes simultaneously. In \ref{fig:AFb}, we plot the spectra of the optimum waveforms for $A = 1$ and $B \in \{1, 40\}$. As can be seen from this figure, with an increase in $B$, in the frequency domain, the waveform shows deep notches and rapid ripples. The reason for this is that,  at large $B$s,  it is required to suppress sidelobes at more delay shifts compared to smaller ones. In the time domain, that means the waveform needs more rapid phase variations which forces the spectrum to be more highly‑rippled. Fig.~\ref{fig:AFc} shows the minimized AF sidelobe energy as a function of the Doppler frequency and the time delay set sizes $A$ and $B$.  Fig.~\ref{fig:AFd} shows $|\chi(d, \nu)|$, defined in \eqref{eq:AFd}, as a function of range and Doppler bins $d$ and $\nu$, respectively. 



In Fig.~\ref{fig:dpc}, we superpose the communication signal on top of  each of the optimum sensing waveforms corresponding to the four target scenarios with clutter at $(r_c = 0.5, d_c = 80m)$.  As can be seen from this figure, through the proposed DPC scheme, we can precancel the interference of the optimum sensing waveform  on the communication signal  irrespective of the target distance and radius. However, given a fixed total transmit power $\P_t = 200 W$ that is shared between sensing and communication tasks, the capacity decreases as we increase the sensing transmit power $\P_s$.

In Fig.~\ref{fig:ofdmdpc}, we superpose the communication signal on the optimum sensing waveform for the case where $A = B = 4$. We consider both WAR and WUR scenarios. Recall that in the WAR scenario,  the communication receiver has full knowledge of the optimum sensing waveform, whereas, in the WUR scenario, the communication receiver has to estimate the sensing waveform. We evaluate the performance of the three superposition schemes introduced in Section~\ref{sec:superposition} denoted by `DPC', `CDMA' and `OFDM'.  Fig. \ref{fig:ofdmdpc} illustrates the MIMO capacity obtained under each scheme as a function of the total transmit power.  We observe that there is only a slight degradation in the  performance of superposition schemes under  the WUR scenario which shows the validity of the proposed WUR decoding approaches. 
It can  also be seen from this figure that the linear transformation (OFDM) approach outperforms the other two schemes. For this plot, we have considered QPSK, (i.e., $K = 4$). Fig.\ref{fig:ofdmISL}, illustrates that, by increasing the order of the modulation $K$, one can further enhance  the capacity but at the cost of a higher ISL, showing a trade-off between sensing and communication performance as a function of $K$.  


%
%

\subsection{ Feasible ISAC  Performance Region}
In this section, we examine the ISAC feasible region. For this purpose, we use minimizing ISL as the design metric and employ DPC as the superposition scheme. In the CC scheme, the optimum sensing waveform is obtained as explained in Section~\ref{sec:ISL}. In the SC scheme where sensing has the higher priority, the communication signal $\vect x^{(c)}$ will be first constructed according to the communication data, and then the sensing signal is formed adaptively with respect to $\vect x^{(c)}$. The optimization of the sensing waveform thus is formulated as
\begin{IEEEeqnarray}{rCl}\label{eq:CFP}
\min_{\vect x^{(s)} } &&\;\;  \xi_{\text{ISL}}(\vect x) \nonumber\\
\text{s.t.} &&\;\;  \sum_{m = 0}^{M_c -1}\sum_{\ell = 0}^{L-1} \mathbb E\left [\| \vect x_{m,\ell}^{(s)}\|^2 \right ]\leq \P_s.
\end{IEEEeqnarray}

Compared with (\ref{eq:CFP}), we observe that the only difference is the missing expectation in the objective function. This is due to the fact that  $\vect x^{(c)}$ is deterministic in the SC case. Similarly to the CC case, the adaptive design of $\vect x^{(s)}$ is again equivalent to minimizing the PSD variance. The corresponding algorithm is very similar to Algorithm \ref{alg:water}, except that we update the power of each symbol of  each subcarrier using $\P_{m,\ell}^{(s)}=(\lambda-\|\vect x_{m,\ell}^{(c)}\|^2)^+$. 

\begin{figure}
  \centering
  \includegraphics[width=0.31\textwidth]{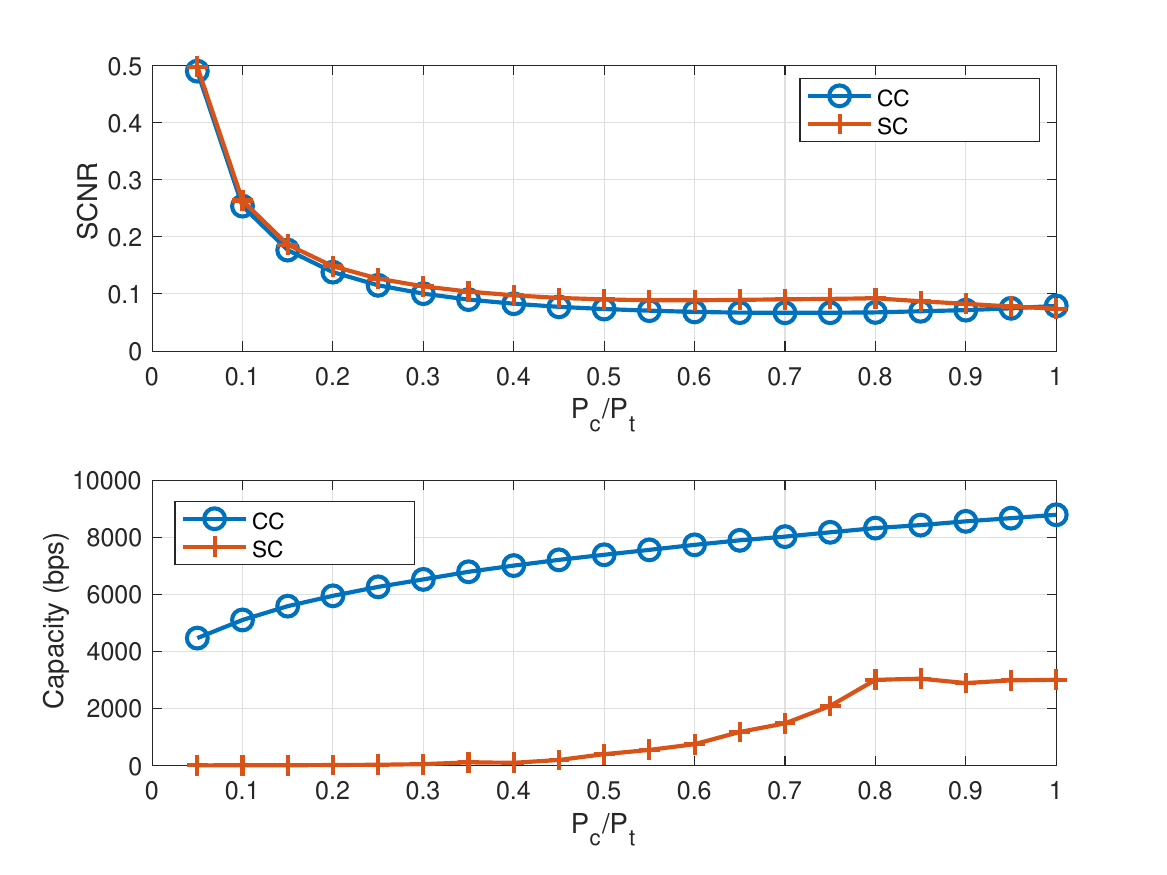}
  \caption{SCNR and communication capacity versus  $\P_c/\P_t$.}\label{fig:perf}
  \vspace{-0.6cm}
\end{figure}

\begin{figure}
  \centering
  \includegraphics[width=0.31\textwidth]{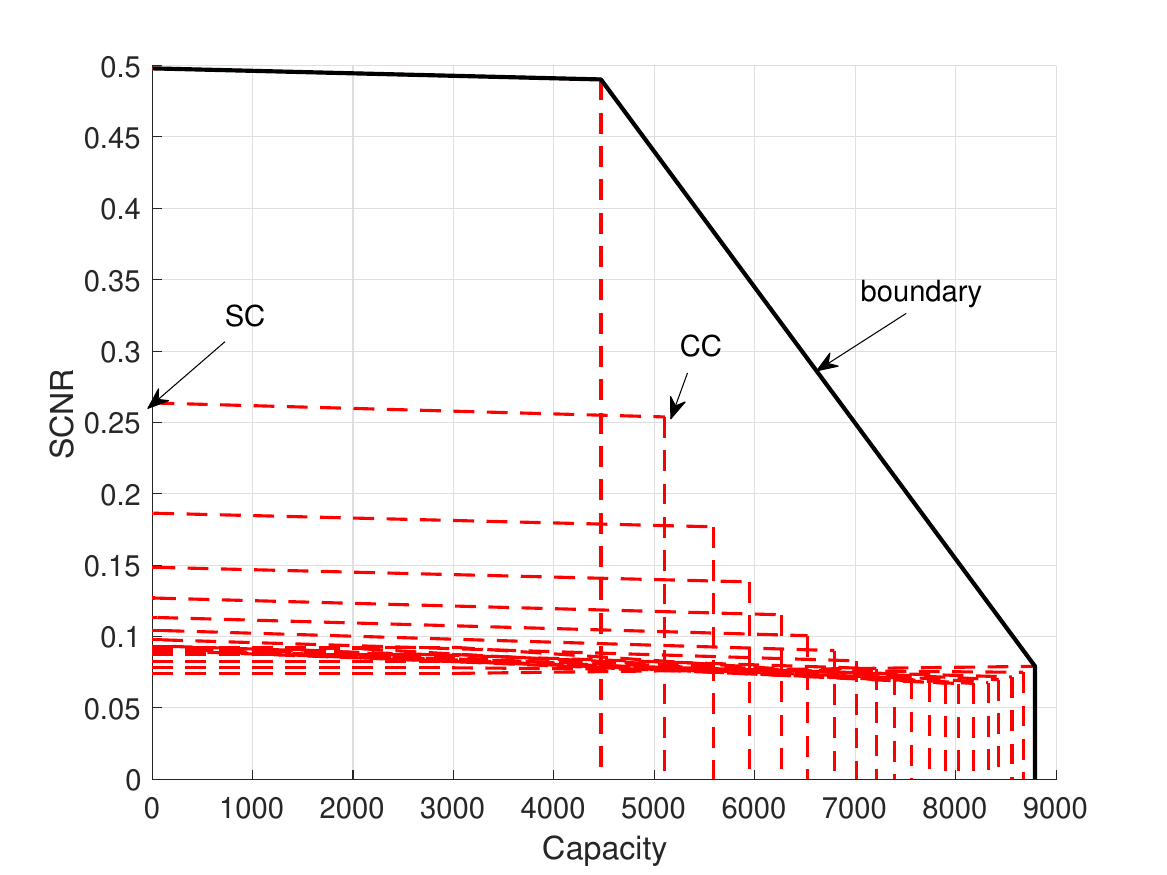}
  \caption{An illustration of ISAC feasible performance region.}\label{fig:region_numerical}
  \vspace{-0.4cm}
\end{figure}

  In our simulation setup, we assume 2048 subcarriers, with starting frequency of 6GHz and frequency spacing of 240kHz (thus the total bandwidth approximately equals 550MHz). We assume that the distances between the ISAC transceiver and target, and between the target and communication receiver, are 40 and 20 meters, respectively, while the line-of-sight (LOS) path between the ISAC transceiver and communication receiver is 50 meters. The pathloss model is assumed to be $48+20\log_{10}d$(dB), where $d$ is the distance in meters. The PSD of the noise is $-194$dBm/Hz, while the total transmit power is 20mW. The reflection coefficients of directions to the ISAC transceiver and communication receiver are assumed to be 1 and 0.2, respectively. Since the two propagation paths result in a frequency-selective channel, the communication power allocated to different subcarriers is obtained from water-filling.
The corresponding SCNR and channel capacity, obtained from the bit error rate of QAM and the assumption of a  binary symmetric channel, are plotted in Fig. \ref{fig:perf}. In this figure, we assume that the proportion of communication power $\P_c/\P_t$ ranges from $\frac{1}{20}$ to 1.  We observe that, in terms of sensing performance (SCNR), the SC scheme is only marginally better than the CC in Algorithm \ref{alg:water}. Meanwhile, in terms of the communication channel capacity, the CC scheme substantially outperforms the SC scheme; in particular, when the proportion of communication signal power is small, the channel capacity of SC is close to 0, which means that the interference from the sensing signal is detrimental. This does not imply that the SC strategy should be discarded, since it has not been optimized. 
Based on the performance metrics in Fig. \ref{fig:perf}, we plot the feasible performance region of ISAC in Fig. \ref{fig:region_numerical}. We observe that the original feasible regions formed by the time sharing of CC and SC construct a non-convex region. The time sharing between the power allocation schemes forms a convex region of performance. The dashed-line regions are the result of increasing $\P_c/\P_t$ from $\frac{1}{20}$ to 1.

\section{Conclusions} \label{sec:conclusions}
Using a broadcast channel model, we have  proposed a unified ISAC framework  in which communication and sensing signals are broadcast to  actual communication users and virtual sensing users.  Within this framework, we have proposed different superposition coding schemes,
for  cases where sensing waveform is known or unknown
to the communication receiver. We numerically evaluated the proposed framework under various sensing and communication metrics. We showed  the effect of clutter, Doppler and delay shifts on the optimum waveform design. We have examined the effect of  the clutterer's distance and size on the SCNR of the target. We also observed that with an increase in the number of Doppler‐shift bins, the optimum waveform shows increased spectral ripples and flatter envelopes. We also compared the performance of DPC, CDMA, and OFDM superposition coding schemes under these metrics. For example, we have observed that the OFDM scheme achieves a higher transmission rate, but at the cost of a higher ISL.

\end{document}